\documentclass{aa}
\pdfoutput=1
\usepackage[utf8]{inputenc}
\usepackage{multirow}
\usepackage{graphicx}
\usepackage{natbib}
\usepackage{xcolor}
\usepackage{soul}

\title{Eruption and propagation of twisted flux ropes from the base of the solar corona to 1 au}
\titlerunning{Simulation of the propagation of a TDm flux rope}

\author{F. Regnault\inst{\ref{ias} \and \ref{cea} \and \ref{ssc}}
\and A. Strugarek\inst{\ref{cea}}
\and M. Janvier\inst{\ref{ias}}
\and F. Auch\`ere\inst{\ref{ias}}
\and N. Lugaz\inst{\ref{ssc}}
\and N. Al-Haddad\inst{\ref{ssc}}
}

\institute{
Universit\'e  Paris-Saclay,  CNRS,  Institut d’Astrophysique Spatiale, 91405 Orsay, France \label{ias}
\and 
D\'epartement d’Astrophysique/AIM, CEA/IRFU, CNRS/INSU, Univ. Paris-Saclay \& Univ. de Paris, 91191 Gif-sur-Yvette, France \label{cea} 
\and Space Science Center, Institute for the Study of Earth, Oceans, and Space, and Department of Physics and
Astronomy, University of New Hampshire, USA \label{ssc}
}

\newcommand{\thinu}{\emph{thin1}}
\newcommand{\thind}{\emph{thin2}}
\newcommand{\thint}{\emph{thin3}}
\newcommand{\thin}{\emph{thin}}

\newcommand{\ie}{\textit{i.e.}}

\newcommand{\thicku}{\emph{thick1}}
\newcommand{\thickd}{\emph{thick2}}
\newcommand{\thickt}{\emph{thick3}}
\newcommand{\thick}{\emph{thick}}

\newcommand{\mbf}[1]{\mathbf{{#1}}}

\newcommand{\aT}{\ensuremath{\alpha_{\mathrm{T}}}}
\newcommand{\aV}{\ensuremath{\alpha_{\mathrm{V}}}}
\newcommand{\aB}{\ensuremath{\alpha_{\mathrm{B}}}}
\newcommand{\an}{\ensuremath{\alpha_{n}}}
\newcommand{\ab}{\ensuremath{\alpha_{\beta}}}
\newcommand{\B}{\ensuremath{B}\xspace}

\newcommand{\Br}{\ensuremath{B_{\mathrm{r}}}}

\newcommand{\Bt}{\ensuremath{B_{\theta}}}

\newcommand{\Bph}{\ensuremath{B_{\varphi}}}

\newcommand{\cm}{cm${}^{-3}$}
\newcommand{\euh}{EUHFORIA}

\newcommand{\V}{\ensuremath{V}}

\newcommand{\Bpe}{\ensuremath{B_{\perp}}\xspace}
\newcommand{\Is}{\ensuremath{I_{\mathrm{s}}}}

\newcommand{\be}{\ensuremath{\beta}\xspace}
\newcommand{\gam}{\ensuremath{\gam}\xspace}
\newcommand{\env}{\ensuremath{\approx}\xspace}

\newcommand{\np}{\ensuremath{n}}
\newcommand{\Vp}{\ensuremath{V}}

\newcommand{\is}{\textit{in situ}}

\newcommand{\ptot}{\ensuremath{p_{\mathrm{tot}}}}
\newcommand{\pth}{\ensuremath{p_{\mathrm{th}}}}
\newcommand{\Rsol}{\ensuremath{R_{\odot}}}

\newcommand{\eg}{\textit{e.g.~}}

\begin{document}

\date{Received 2022-07-12 / Accepted 2022-10-27}

\abstract{
Interplanetary Coronal Mass Ejections (ICMEs) originate from the eruption of complex magnetic structures occurring in our star’s atmosphere. Determining the general properties of ICMEs and the physical processes at the heart of their interactions with the solar wind is a hard task, in particular using only unidimensional \is\ profiles. Thus, these phenomena are still not well understood. 
}{
In this study we simulate the propagation of a set of flux ropes in order to understand some of the physical processes occurring during the propagation of an ICME such as their growth or their rotation. 
}{
We present simulations of the propagation of a set of flux ropes in a simplified solar wind. We consider different magnetic field strengths and sizes at the initiation of the eruption, and characterize their influence on the properties of the flux ropes during their propagation. We use the 3D MHD module of the PLUTO code on an Adaptive Mesh Refinement grid.
}{
The evolution of the magnetic field of the flux rope during the propagation matches evolution law deduced from \is\ observations. We also simulate \is\ profiles that spacecraft would have measured at the Earth, and we compare with the results of statistical studies. We find a good match between simulated \is\ profiles and typical profiles obtained in these studies. During their propagation, flux ropes interact with the magnetic field of the wind but still show realistic signatures of ICMEs when analyzed with synthetic satellite crossings. We also show that flux ropes with different shapes and orientations can lead to similar unidimensional crossings. This warrants some care when extracting magnetic topology of ICMEs using unidimensional crossings.
}{}
\maketitle

\section{Introduction}
\label{sec:intro}

An interplanetary coronal mass ejection (hereafter, ICME) corresponds to the magnetically dominated plasma that is the interplanetary counterpart of eruptions occurring in the solar corona. 
When these magnetic structures (called magnetic ejecta, ME) are expelled from the solar corona, they propagate in the interplanetary medium. During the propagation, ICMEs and their different substructures can be probed by \is\ instruments on board spacecraft like ACE \citep{chiu1998}, Wind \citep{harten1995} situated at the first Lagrange point (L1), or like Parker Solar Probe \citep{fox2016} and Solar Orbiter \citep{muller2020} which span a wide range of heliospheric distances or finally like STEREO A and B \citep{kaiser2007} that sample other heliographic longitudes than observatories at L1.

If ICMEs are fast enough (\ie\ faster than the local solar wind), they may accumulate solar wind material ahead of them and form a sheath. 
In that case, the solar wind encountered during the propagation of the ICME does not have enough time to flow around the ICME. It thus creates a region of heated and compressed solar material. Moreover, the expansion of the ME during its propagation enhances the accumulation of the solar wind material \citep{siscoe2008}. Recent studies \citep{moissard2019,kilpua2020} have shown that this region is more turbulent than the local solar wind by computing the power spectrum of the magnetic field fluctuations and deducing its anisotropy in the sheath and in the solar wind. They found that the orientation of the magnetic field fluctuations and their magnitudes are more erratic in the sheath than in the solar wind.
However, ICMEs do not always have an observable sheath. In fact, no sheath ICMEs are comparable to the slow ICMEs with a sheath \citep{regnault2020}.

If the speed of the ICME is faster than the local fast mode speed then they can develop a fast-forward shock ahead of the sheath. All ICMES do not have a shock in front of their sheaths. \cite{salman2020a} studied the difference between ICMEs sheaths that has a front shock with those who don't. They found that the magnetic and plasma properties of sheaths with a front shock are higher in magnitude than for those without it.
Moreover, all shocks detected in \is\ data are not necessarily ICME-driven but they can be related to a Stream Interacting Region (SIR) \citep{kilpua2015} which is formed by the interaction of a fast solar wind stream with the slow solar wind stream.

Since it is a magnetically dominated plasma, the ME is characterized by a low plasma $\beta$ (thermal pressure over magnetic pressure), typically $\approx 0.1$. We also observe a decrease of the proton temperature and density compared to the sheath area due to the expansion of the ME along its propagation. This expansion is mainly due to the decrease of the solar wind pressure with the distance \citep{demoulin2009}. The radial expansion of the ME produces a monotonic decrease of the speed profile \citep{klein1982,farrugia1993}, although this decrease is not always present (\eg\ in the case of a fast stream or another ICME overtaking the primary structure,  see more details in \citealt{gulisano2010}). The fluctuations of the magnetic field divided by its magnitude is lower \citep{masias-meza2016} because MEs are more coherent magnetic structures than the solar wind, so the magnetic field is smoother.

By looking at the components of the magnetic field inside the ME, we sometimes observe a smooth rotation of one of the components of the magnetic field. If the latter is observed along with a low proton temperature and a high magnetic field intensity, then the ME is called a magnetic cloud \citep[MC,][]{burlaga1981}. 
This smooth rotation of one component of the magnetic field is interpreted as a trace of the flux rope (hereafter, FR) structure of the MC. Such a shape is often used to describe the magnetic structure of ICMEs. For example, the \cite{lundquist1951} model assumes a flux rope structure with a circular cross-section and an invariant axis. 

One third of MEs are detected as MCs \citep{richardson2004a,wu2011}, although this low proportion may be directly linked to the crossing of the structure.
\cite{kilpua2011} have shown that the observation of the FR signature is highly impacted by the trajectory of the spacecraft through the ICME. Using multi-spacecraft observations of ICMEs, \cite{jian2006,demoulin2013,zhang2013} also found that the trajectory of the spacecraft through the ICME is important for the observation of MC signatures.

We see here one limitation of \is\ data analysis for the study of the property of the propagating ICME. Indeed, the 1D crossing of the 3D complex structure of the ICME allows only to get limited information about it \citep{riley2004a,al-haddad2019a}. Some reconstruction techniques are able to get the global properties of the ICME from the 1D crossings, but it comes with the price of significant approximation of, for example, the magnetic structure (force-freeness, cylindrical symmetry, etc.) following for example the \cite{lundquist1951} or \cite{gold1960} models. For example, the cylindrical symmetry assumption is likely to be far from ideal. According to \cite{riley2004b}, a force-free flux rope should not keep a circular cross-section during its propagation.

Moreover, studying the evolution of the properties of ICMEs is a hard task due to the low number of spacecraft in the interplanetary medium. Ideally, one would need to be able to measure the properties of the same ICME at different distances from the Sun. These ICMEs are rare and are mostly studied in single case studies (see for example \citealt{kilpua2011} for a review of multispacecraft observations with the STEREO mission). \cite{janvier2019} managed to perform a statistical study on ICMEs detected close to Mercury with the Messenger probe, to Venus with the Venus Express probe or close to the Earth with the ACE probe. However, Messenger and Venus Express don't have any \is\ measurements of the properties of the interplanetary plasma (temperature, speed and proton density for example) and it thus limits the conclusions drawn by the paper to the magnetic properties of ICMEs.

Simulations provide an interesting complementary approach to the study of the propagation of a flux rope in the interplanetary medium. In a simulation, one can perform a 3D analysis of the ICME structure. This allows a more detailed analysis of the structure of the ICME but also of physical processes happening during the propagation of a flux rope magnetic structure in a simulated solar wind. We can then compare properties of the simulated ICMEs with properties of actual ICMEs for example by doing synthetic crossings through the ICME and compare with observations.

Simulations of ICMEs have already been performed. However, most of them focus either on the trigger part of the ICME (see \eg\ \citealt{aulanier2010,kliem2012}) or on the propagation of the ICME up to the Earth. The EUHFORIA simulations insert a spheromak magnetic structure at 0.1 au and study its propagation up to the Earth \citep{verbeke2019}. \cite{shen2014} performed a Sun to Earth simulation of the propagation of a magnetic structure that is initiated closer to the Sun but not anchored to it. Inserting the magnetic structure at a certain distance from the Sun can't properly take into account the different physical mechanisms linked to the eruption of the flux rope (for example, its impact on the rotation).

\cite{torok2018} took into account the effect of the eruption mechanism on the properties of the propagating flux rope when performing a Sun-to-Earth simulation of an eruptive prominence event. They did a relaxation of a flux rope structure (anchored at the solar surface), triggered its eruption with converging flows in the photosphere and then studied its propagation up to the Earth.

Similarly, this paper aims mainly at investigating the propagation of a flux rope that erupted within the same simulation domain.
We run a parametric study in order to highlight the physical mechanisms that happen during the propagation of a flux rope in the solar wind.

In Section \ref{sec:setup}, we describe the numerical setup and the flux rope model used in this study. In Section \ref{sec:evol}, we investigate the evolution of the magnetic structure during its propagation. In Section \ref{sec:obs}, we compare the results of the simulations with results obtained with \is\ measurements.
The results are discussed in Section \ref{sec:discussion}.

\section{Numerical setup}
\label{sec:setup}

\subsection{The PLUTO code}
\label{sec:PlutoEqs}

In this study, we use the PLUTO code to solve the 3D MHD equations in an
Adaptive Mesh Refinement (AMR) spherical grid \citep{mignone2012}. PLUTO is a
multiphysics and multisolver code using a finite-volume Godunov type scheme. 
Such a scheme requires computing the flux through the interfaces of each elementary 
volume paving the simulation domain. To do so, we use the Harteen, Lax, van Leer solver \citep{toro2009}.

The MHD equations are written in their conservative form: 
\begin{eqnarray}
    \frac{\partial\rho}{\partial t} + \nabla \cdot (\rho \mathbf{v}) &=& 0,\\
    \frac{\partial\rho\mbf{v}}{\partial t} + \nabla \cdot [\rho\mbf{v}\mbf{v} -
    \mbf{B} \mbf{B}] + \nabla \ptot &=&  \rho \mbf{g},\\
    \frac{\partial \mathcal{E}}{\partial t} +
    \nabla \cdot [(\mathcal{E} + \ptot)\mbf{v} -(\mbf{v}.\mbf{B})\mbf{B})] &=&
     \rho \mbf{v}.\mbf{g},\\
    \frac{\partial \mbf{B}}{\partial t} - \nabla \times ( \mbf{v} \times \mbf{B} )
     &=& 0 \label{eq:induction}
    \label{eq:MHD}
\end{eqnarray}

where $\mbf{B}$ is the magnetic field intensity, $\rho$ the mass density, $\mbf{v}$ the
speed, $\mbf{g}$ the gravitational acceleration. $\ptot = \pth + \mbf{B}^2/2$
corresponds to the total pressure (thermal + magnetic). Finally, $\mathcal{E}$ is
the total energy density which is given by :

\begin{eqnarray}
    \mathcal{E} = \rho \epsilon + \frac{1}{2} \rho \textbf{v}^2 + \frac{1}{2} \textbf{B}^2
    \label{eq:Etot}
\end{eqnarray}

To close the system of equations, we use an ideal equation of state for $\epsilon$ :

\begin{eqnarray}
    \rho \epsilon = \frac{p_{\mathrm{th}}}{(\gamma - 1)} \label{eq:eos-poly}
\end{eqnarray}
The ratio of specific heats, $\gamma$, is set to 1.05 to maintain a heated corona. This value is chosen to heat enough the low corona so that it can drive a solar wind (see Section \ref{sec:solarwind} for more details).

Equations are solved in a spherical geometry ($r$,$\theta$,$\varphi$) in a grid
that ranges from 1 to 420 \Rsol\ for $r$, from 0 to $\pi$ for $\theta$ and
from 0 to $2\pi$ for $\varphi$. The initial grid is stretched in the $r$ direction so that the spatial resolution decreases while going away from the Sun following a logarithm. Other simulations \citep{torok2018,poedts2020} use a non-uniform grid in $r$ with a higher spatial resolution closer to the Sun.
Moreover, the AMR grid adapts its refinement in function of the gradient of the total energy
density. This allows us to keep a finer grid in the region of the propagating ICME while having a coarser grid far from it. 
We use four levels of AMR with a refinement ratio set to 2. At maximum spatial resolution (level = 4), the step in $\theta$ and $\varphi$ is 0.012 rad and $\env 0.012$ \Rsol\ in the $r$ direction close to the Sun and $\env 2.5$ \Rsol\ 
at 210 \Rsol.

We do not prescribe any resistivity in this model. However, these ideal MHD equations can still lead to a change of 
connectivity of magnetic field lines due to numerical resistivity. 

\begin{figure*}
    \centering
    \includegraphics[width=1\linewidth]{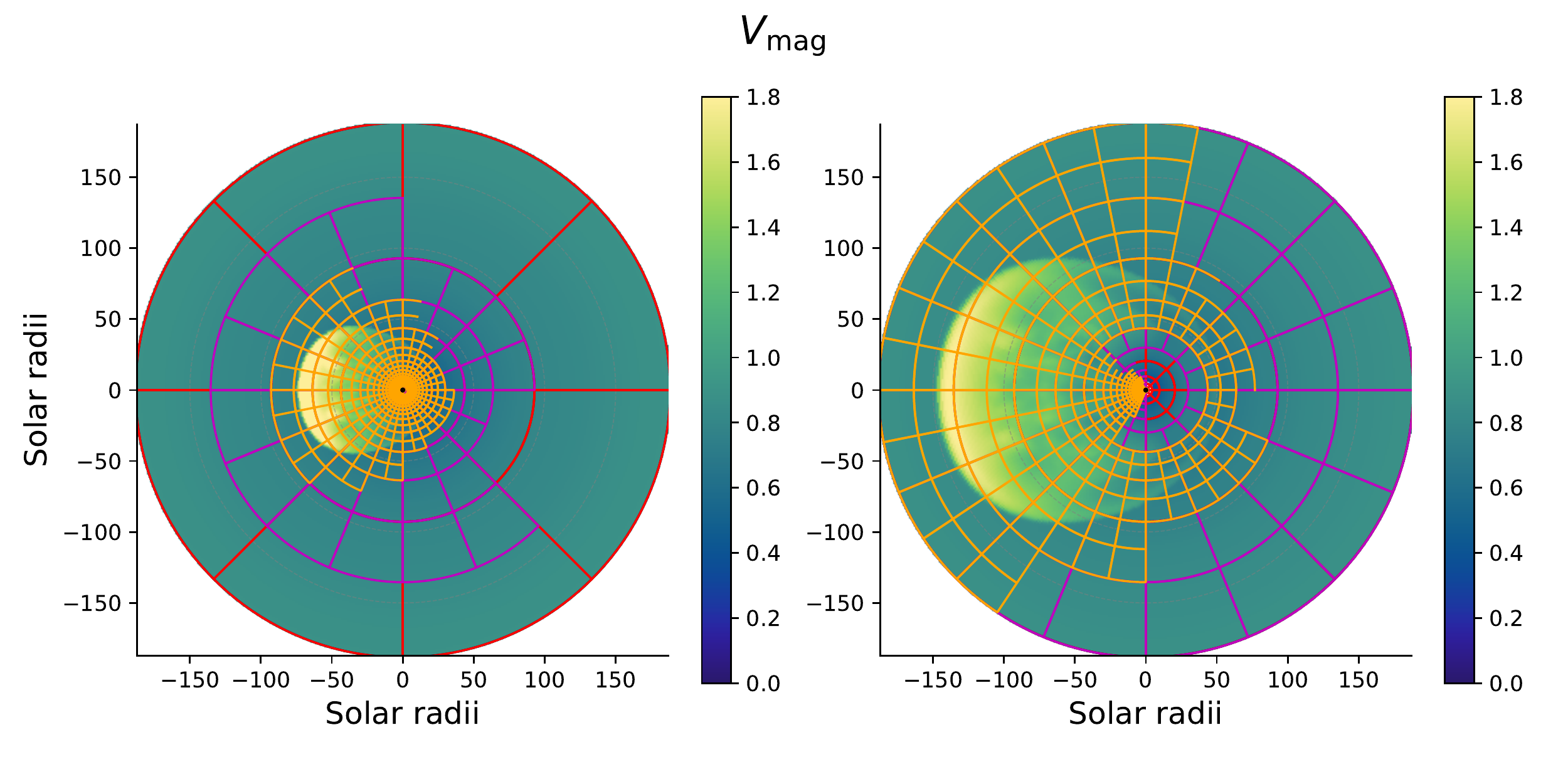}
    \caption{Snapshots of the equatorial slice of the PLUTO simulation with the propagating TDm flux rope. The gradient of color shows the velocity magnitude in PLUTO units (1 $\env$ 438 km/s). Orange areas correspond to level = 4 of the AMR grid,  purple to level = 3, red to level = 2.
The level = 1 does not appear here.}
    \label{fig:AMR_examples}
\end{figure*}

Figure \ref{fig:AMR_examples} shows two equatorial slices of the
speed magnitude at different times during the propagation of a simulated ICME. Boxes with
different colors show the level of refinement in the equatorial slice. We see
here that the high resolution (in orange) is following well the ICME, shown in yellow, during its propagation
in the heliosphere.

\subsection{Solar wind model}
\label{sec:solarwind}

We use the solar wind model of \cite{reville2015}, initialized
using the Parker model \citep{parker1958} with a polytropic equation of state :

\begin{equation}
P \propto \rho^{\gamma}
\end{equation}

with $P$ the pressure, $\rho$ the density and $\gamma = 1.05$ the ratio of
specific heat. This assumption produces a polytropic heating of the full corona
and makes it almost isothermal at $\approx 10^6$ K.
The surface magnetic field is set up to be a dipole with a maximum field strength at the surface of $2.9$ G. This
configuration is close to the magnetic configuration of the Sun during the solar minimum
\citep{reville2017}. The polytropic assumption is used only during the initialization, the model then uses an ideal 
equation of state (see \S \ref{sec:PlutoEqs}). Such a model drives a solar wind that has the same speed in all direction, we thus calibrate the model such that we have a speed of $\approx 440$ km/s at 1 au. 
We also pay attention to the mass loss rate and the angular
momentum loss of the simulated solar wind. The model is calibrated so that the mass loss rate and the angular momentum (averaged from 3 to 250 \Rsol) correspond to the ones measured in the solar system (see, \eg\ \cite{finley2018} for the angular momentum and \cite{schwadron2008} for the mass loss rate).
For more details about the solar wind model, see \cite{reville2015}.

Once the simulation evolves in time, the hydrodynamic properties (speed, density) and
the magnetic properties will interact with each other and produce a self-consistent MHD solar
wind.

\subsection{Flux rope model}

\begin{figure}
    \centering
    \includegraphics[width=\linewidth]{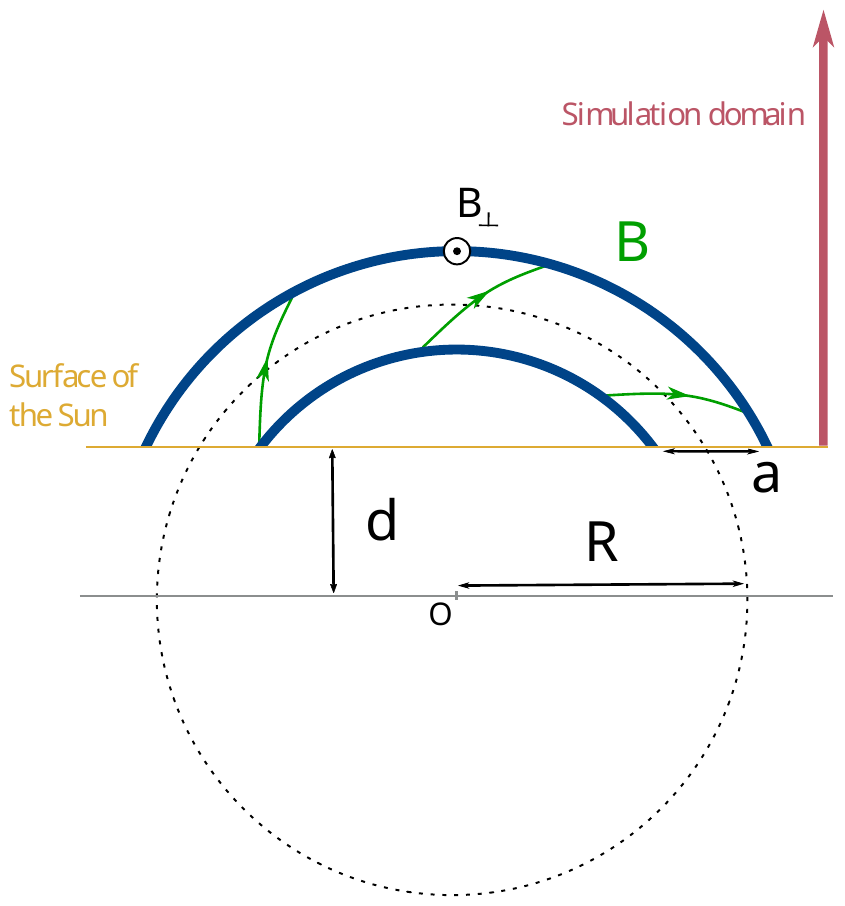}
    \caption{2D schematic of the TDm setup. The yellow line corresponds to the inner boundary of the simulation domain and the blue line corresponds to the edge of the flux rope of minor radius $a$ and major radius $R$. The $d$ parameter controls how much the flux rope is buried in the solar surface \citep{titov2014}. The magnetic field of the twisted flux rope appears in green. \Bpe\ correspond to the ambient magnetic field.}
    \label{fig:TDm_scheme}
\end{figure}

In this study, we use the analytical formulation of a flux rope described in
\cite{titov2014}. The modified Titov-Démoulin (hereafter, TDm) flux rope provides
an analytical formulation of a ring current that has a magnetic flux rope
structure. The model is built so that the magnetic structure is force-free if the ambient
magnetic field is perpendicular to the torus plane and distributed
uniformly along its toroidal segment. We call this magnetic field $B_{\perp}$. A horizontal magnetic field
can meet these requirements, but a bipolar magnetic configuration allows us to have a magnetic field 
that is approximately uniform along the toroidal segment. Figure \ref{fig:TDm_scheme} shows a schematic of the
TDm flux rope with the different parameters for the TDm model.  $R$ and $a$ are
the major and minor radius of the torus, respectively. $d$ is the depth at
which the torus is buried below the solar surface represented by the ($x$,$y$)
plane in this figure.

In order to be force-free, the intensity of the ring current should
be equal to the Shafranov current \Is\ \citep{shafranov1966} which is given by
\begin{eqnarray} 
    I_{\mathrm{s}} \approx - \frac{4 \pi R B_{\perp} /
\mu}{\ln\left(\frac{8R}{a} \right) - \frac{3}{2} + \frac{l_i}{2}},
\end{eqnarray}

with $\mu$ the magnetic permeability and $l_i$ is the internal self-inductance 
per unit length of the rope that is $\approx 1$ in this case.

In this work, we study the effect of the initial magnetic field on the
properties of the propagating flux rope. To do so we define the
$\zeta$ parameter such as : 
\begin{equation}
    I = \zeta I_{\mathrm{s}}.
\end{equation}
with $I$ the intensity of the ring current of the flux rope.

The $\zeta$ parameter allows us to control the initial magnetic field of the flux
rope. It thus means that a flux rope initialized with $\zeta \neq 1$ is not in a force free state. 
The TDm flux ropes in this study are initialized out of
equilibrium. This has the advantage that we skip the costly relaxation step and immediately simulate the eruptive phase. A more thorough treatment of the initial relaxation phase is deferred to future works.

The original TDm model was proposed by \cite{titov2014} in two flavors.
The difference between the two cases lies in the
current density distribution inside the torus. Case 1 has a "hollow-core"
current distribution, meaning that the twist
of the magnetic field, thus the current, is concentrated in the outer edge of the torus. Such current distribution has been observed in MCs by \cite{lanabere2020}. Indeed, they studied the distribution of twist in FR observed at 1 au (assuming a Lundquist model) and they found a constant twist inside the core of the FR and a twist a higher by a factor 2 at the boundary.
Case 2 has a
parabolic distribution of the current, meaning that the magnetic twist is
distributed over the whole cross-section of the flux rope. In this study, we only use the Case 1 flux ropes, leaving the propagation of the Case 2 flux ropes for another study. 

The code developed to insert a TDm flux rope in the PLUTO data files has been released and is available online\footnote{\url{https://github.com/fregnault/pyTDm}}.

\begin{figure*}
    \centering
    \includegraphics[width=\linewidth]{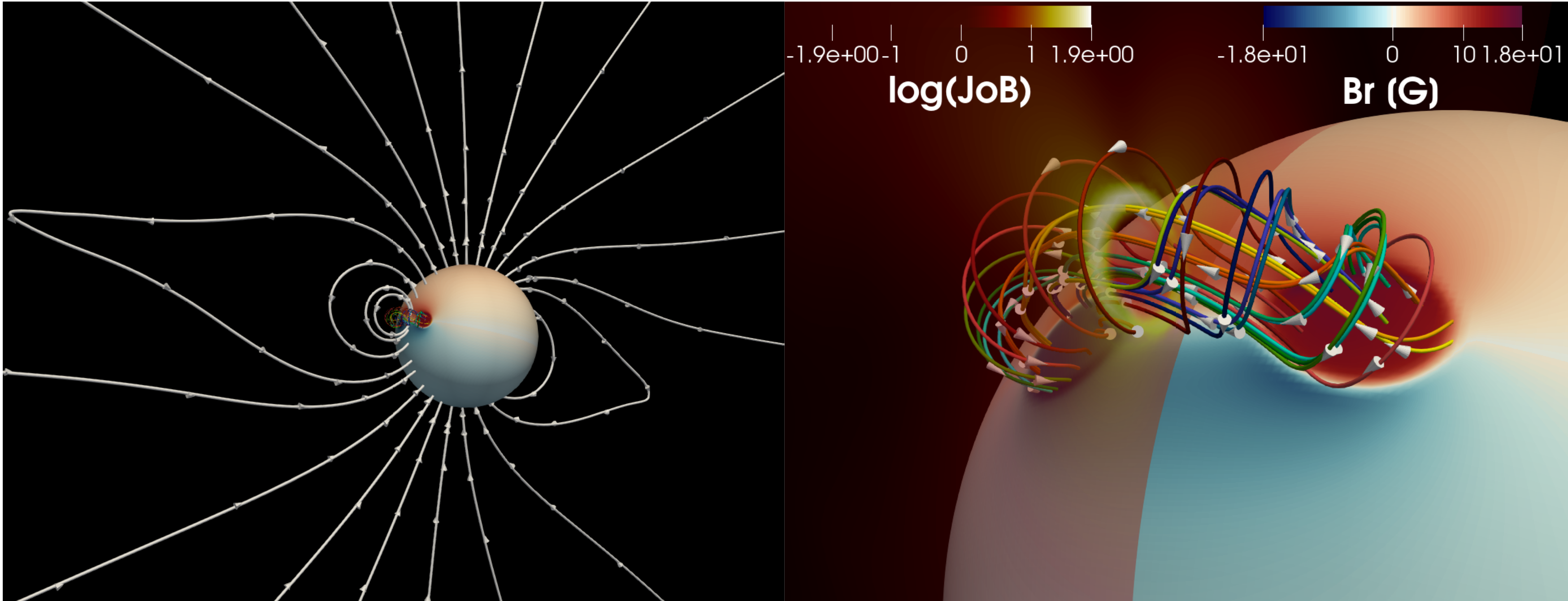}
    \caption{Two visualizations of the initiation of the TDm flux rope in the solar wind model. The left panel shows the global magnetic field of the solar wind with white streamlines. The sphere corresponds to the inner limit of the simulation domain $r = \Rsol$ and the colors correspond to $Br$. Blue means $\Br <0$ and red $\Br > 0$. The right panel shows a zoom on the TDm flux rope. The colored streamlines correspond to the magnetic field lines of the flux rope starting from a circle of a radius $0.9a$ chosen at the footpoints of the flux rope in the negative polarity. The transparent red-yellow slice shows $\log(\frac{J}{B})$ (with $J = \nabla \times B$ the current density) in the plane perpendicular to the torus plane. The yellow ring shows the annular distribution of the current specific to the case 1 of the TDm.}
    \label{fig:TDm_setup}
\end{figure*}

Figure \ref{fig:TDm_setup} shows an example of a TDm initialization in the global wind model. On the
right side of the figure, a spherical slice shows \Br\ (the blue color
corresponds to negative \Br). The colored lines correspond to magnetic field
lines chosen in the negative polarity, and anchored in a 0.9$a$ width circle at the center of the flux rope.
The shape of the ring current that appears in yellow/white is typical of
the Case 1 as mentioned earlier. On the left side, a zoomed-out view of
the global solar wind magnetic field is shown with white streamlines. One can
see here that the TDm flux rope is initialized within the equatorial streamer
region where the density is higher and the magnetic field is in a closed
configuration.

\subsection{Boundary conditions}

The 
simulation domain is surrounded by two layers of cells (called ``ghost cells'') which values set the behavior at 
the boundary.
The outer boundary conditions ($r = 420 \Rsol$) are the Neumann boundary conditions, meaning a constant gradient.
At the poles ($\theta = 0$ and $\pi$) the conditions are axisymmetric. It means that the components parallel 
to the $\theta$ direction change sign and the others remain unchanged. At the $\varphi = 0$ and $2\pi$ boundaries, 
the 
condition is also of Neumann type which is enough for the asymmetric solar wind. ICMEs simulated in this study 
propagate in the direction radially opposite of this boundary to avoid any spurious effect due to its presence. 
The inner boundary condition ($r = \Rsol$) is set differently close to the flux rope and far away from it. We define the ratio 

\begin{equation}
\chi_{\rm BC} = \frac{P_{\rm mag}^{\rm TDm}}{P_{\rm mag}^{\rm SW}},
\end{equation}

with $P^{\rm SW}_{\rm mag}$ and $P^{\rm TDm}_{\rm mag}$ the magnetic pressure of the solar wind (before the insertion of the TDm) and of the TDm flux rope only respectively. 

Cells with $\chi_{\rm BC} < 0.5$ are far from the flux rope and the boundary conditions are the ones of the classic
solar wind model. In that case, the poloidal ($r$ and $\theta$) speed is set to be parallel to the poloidal magnetic
field to limit the production of current and ensuring that the Sun is a perfect conductor.

Cells with $\chi_{\rm BC} > 0.5$ are close to the flux rope and the boundary conditions are set such that there is a 
fixed gradient for all the components of the magnetic field to extend the magnetic field lines in the two layers of 
ghosts cells. We apply a reflective condition on the $r$ component on the speed, simulating a very dense solar 
surface. The $\theta$ component of the velocity has a Neumann type boundary condition at this boundary.

The \thick\ flux ropes (see Section \ref{sec:param_study}) have an Neumann type inner boundary condition for the pressure while the \thin\ ones have a condition on the pressure that ensures hydrostatic 
equilibrium in the ghost cells. In the end, we find that this condition has a low impact on the property of the flux rope. 

\subsection{Flux rope properties for the parametric study}
\label{sec:param_study}

To study the effects of the flux rope properties on its propagation, we investigate in the following the influence of the initial magnetic field strength and thickness of the flux rope. We
initialize six different TDm flux ropes as follows: we choose two sets of TDm flux ropes, with one set corresponding to a thickness $a$ of $0.1 \Rsol$ and the other to a thickness of $0.05 \Rsol$.

Each of these sets then has three different initial
magnetic fields, for which the parameters are summarized in Table \ref{tab:cases}. All are initialized at the same location on the Sun's surface, at ($\theta_0$, $\varphi_0$) = ($\frac{\pi}{2}$,$\pi$) aligned along the solar
equator. 
The $\Delta$ parameter (on the last column of Table \ref{tab:cases}) controls
the width of the ring current (see the yellow ring in the left panel of Fig. \ref{fig:TDm_setup}), it is set to be $\frac{a}{10}$ so that the
condition $\delta \ll 1$ is met (assumption used for the analytic development of
the TDm, see \citealt{titov2014}). All six flux ropes have a major radius $R=0.3$ \Rsol\ and have $d = 0.15$ \Rsol.

The
different initial magnetic fields range from 16 G to 84 G, which corresponds 
typically to the magnetic field observed in prominences \citep{casini2003}.
The magnetic flux of the flux ropes ranges from $1.3$ to $4.4~\times~10^{21}$ Mx that
is close to the flux of large sunspots \citep{zwaan1987}.
Therefore, the six flux ropes have realistic, solar-like values of magnetic field intensity and flux. Although one may argue that the largest prominences are generally quiescent sun regions rather than large sunspots ones.

\begin{table}[h]
\begin{center}
    \begin{tabular}{c c c c c c c c c c}
    \hline
    Case & $B_{\mathrm{FR}}$ [G]  & $F$ [$10^{21}$ Mx] & a [\Rsol] &
    $\Delta$ [\Rsol]\\
    \hline
    \thicku\ & 16 & 2.2 & \multirow{3}{*}{0.1}  & \multirow{3}{*}{0.01} \\
    \thickd\ &  22 & 3.1 &  & \\
    \thickt\ &  31 & 4.4 &  & \\
   \hline
    \thinu\ &  34 & 1.3 & \multirow{3}{*}{0.05}   & \multirow{3}{*}{0.005}\\
    \thind\ &  48 & 1.8 &   &  \\
    \thint\ &  84 & 3.3 &   &  \\
    \hline
\end{tabular}

\caption{Flux rope properties for the 6 different
initial TDm. From left to right : Name of the case, magnetic field intensity
at the core of the initial flux rope in G, magnetic flux of the flux rope in units of $10^{21}$ Mx, minor
radius in \Rsol\ of the torus, and value of the $\Delta$ parameter.}
\label{tab:cases}
\end{center}
\end{table}

\section{Evolution of the magnetic structure of the simulated ICMEs}
\label{sec:evol}


\subsection{Eruption}
\label{sec:erup}

\begin{figure*}
    \centering
    \includegraphics[width=\linewidth]{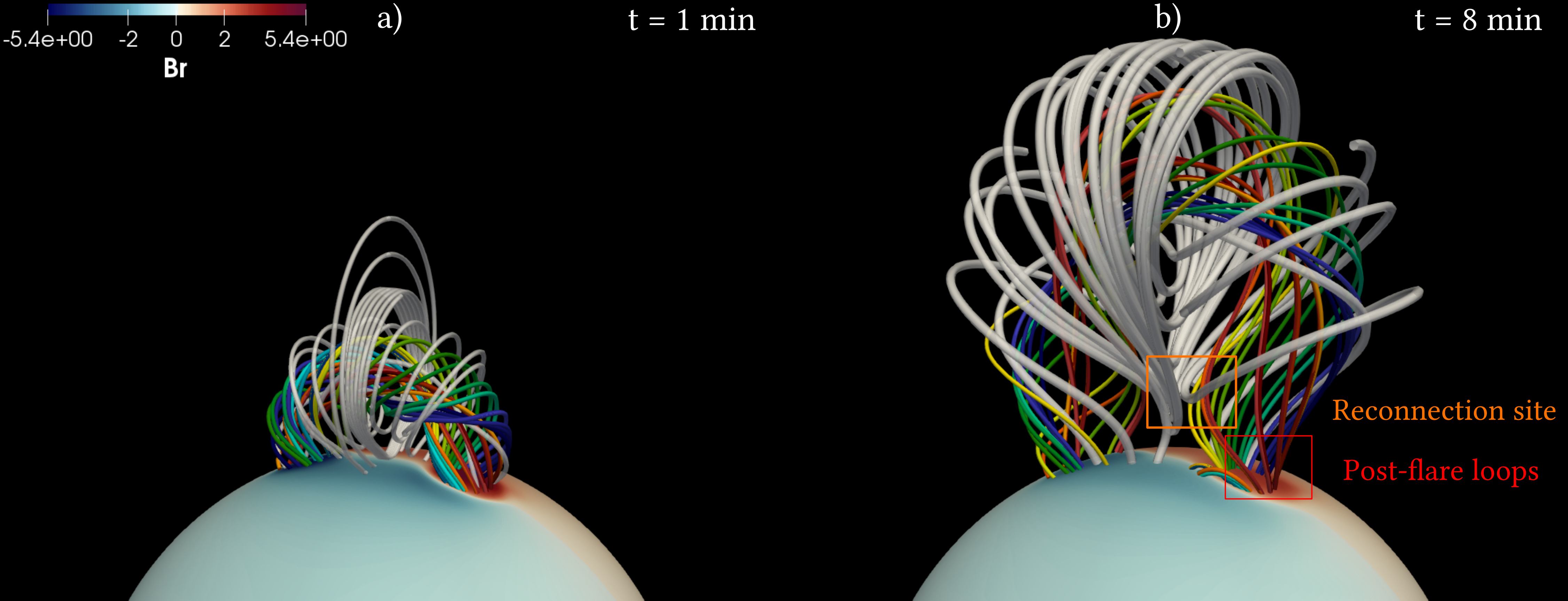}
    \caption{Visualization of the eruption of the \thickd\ flux rope. The spherical slice and the streamlines color are the same as in Figure \ref{fig:TDm_setup}. Here, the white streamlines start from the $\theta = \frac{\pi}{2}$, $\varphi = \pi$ line along the radial direction. Panel a) shows the flux rope 1 minute (physical time) after its eruption and the panel b) 8 minutes after.}
    \label{fig:TDm_erup}
\end{figure*}

In this study, the flux ropes are initialized in an unstable state and erupt quickly after their insertion.
Figure \ref{fig:TDm_erup} shows 3D visualizations of the TDm flux rope 1 (left panel) and 8 minutes (physical time) after the insertion of the flux rope structure for the \thickd\ case.
We would like to remind that the speed gained by the FR comes from Lorentz forces (no ad-hoc speed is being added to the simulation) and that the initial FR has magnetic field and flux similar to that of magnetic structure found in the solar atmosphere. However, it evolves in the medium that did not have time to relax due to the insertion of the FR which is unstable right from the beginning.

The colored streamlines 
correspond to the magnetic field lines of the flux rope starting from a circle of a radius $0.9a$ at the negative polarity where we find the footpoints of the flux rope.
On the right panel, 8 minutes after the insertion, the \thickd\ FR has a speed of $\approx 900$ km/s at 1.5 \Rsol. 
For comparison, the 
\thint\ FR has a speed of $\approx 4400$ km/s  and reaches 1.5 \Rsol\ 1.5 minutes after its insertion. This case is the more magnetized (see Table \ref{tab:cases}) and thus produces the largest initial Lorentz forces propelling the eruption. According to \cite{chen2011}, plane-of-sky CME speeds can occasionally reach 3500 km/s. The \thint\ case therefore corresponds to an extreme case in 
terms of speed during the eruption.

During the eruption, some field lines change their connectivity, with some becoming the low-lying loops shown in the red rectangle. These loops are reminiscent of the so-called post-flare loops, and are 
formed by reconnection happening during the eruption of the flux rope in the trailing region indicated with the orange rectangle. This is expected according to the standard model of solar eruptive flares in 2D (\citealt{carmichael1964,sturrock1966,hirayama1974,kopp1976} type) or in 3D 
(see \citealt{janvier2017} and references therein).
This kind of structure resemble loops that are observed in ultraviolet images after an eruption (\citealt{aulanier2012} and references therein, \eg\ \citealt{su2006,warren2011}).
We note however that the reconnection happening in these simulations is numerically driven, since we don't have any explicit resistivity in the simulation.

During the early propagation of the \thick\ flux ropes, we observe two high speed flows starting from the inner boundary of the simulation. These high speed streams cover a small area in the wake of the ICME and reach typical speeds of the bulk of the ICME itself. Because of the high speed, a region of low pressure (10 orders of magnitude smaller than the solar wind pressure) is formed. A threshold beyond which the pressure value is set to $10^{-12}$ (in PLUTO units) for the cells that encountered this situation. Interestingly, \thin\ flux ropes  have almost no high speed streams in their wake.

\subsection{Propagation}
\label{sec:prop}

\begin{figure*}
    \centering
    \includegraphics[width=\linewidth]{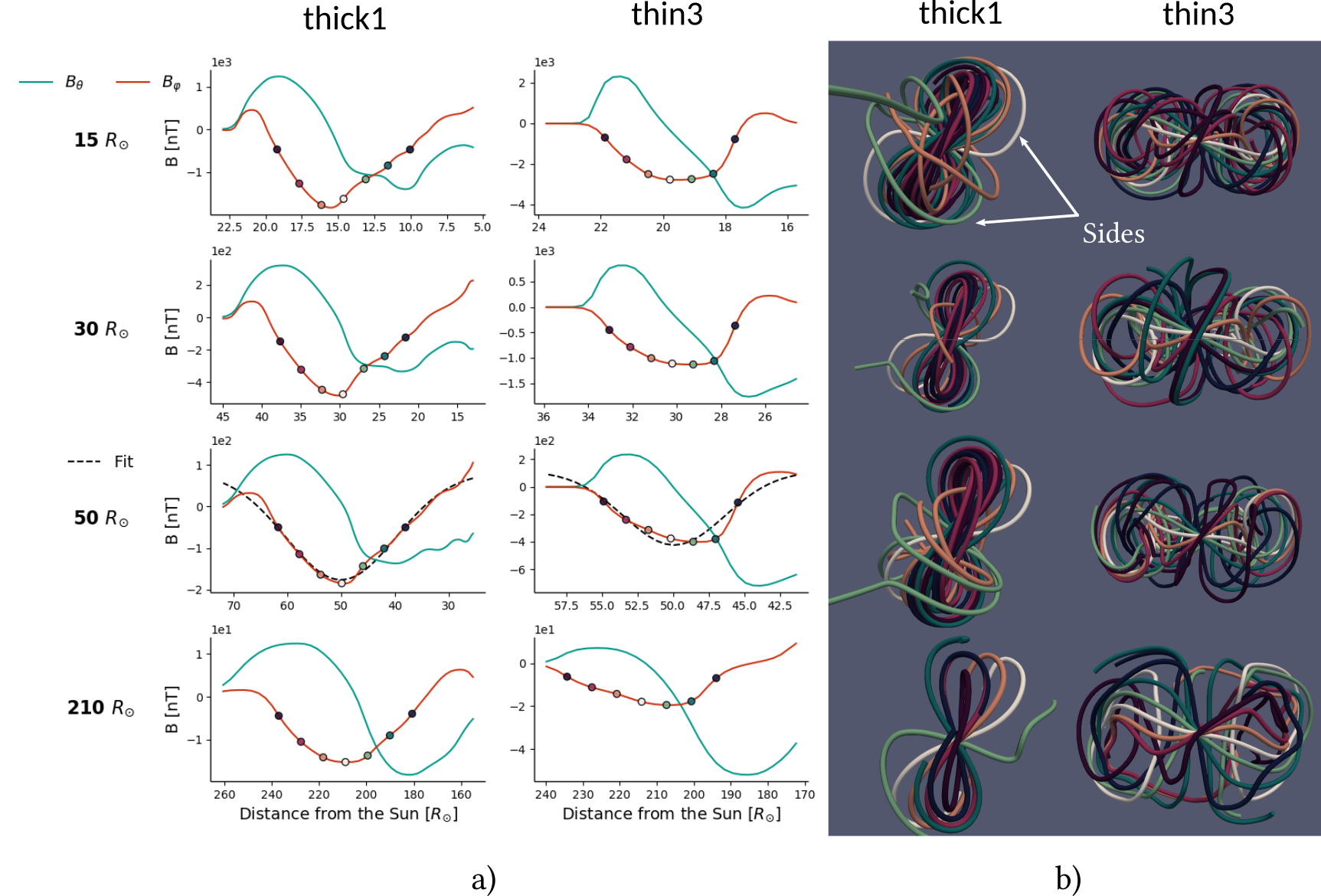}
    \caption{Panel a) shows 1D cuts of \Bt\ (in green) and \Bph\ (in red) at the nose of the \thicku\ (left) and \thint\ (right) flux rope at 15, 30, 50 and 210 \Rsol\ from top to bottom. The black dashed line for each FR at 50 \Rsol\ is a Gaussian fit of \Bph\ (see Section \ref{sec:evol_param}).  The dots show the position of the seed streamline displayed in panel b) (of the corresponding color) projected on the \Bph\ profile. The $x$-axis is decreasing from left to right in order to compare with Figures \ref{fig:SC_thick} and \ref{fig:SC_thin}. Panel b) shows the magnetic structure of the \thicku\ (left) and \thint\ (right) cases. All streamlines originate from seeds that are on the ($\theta = \frac{\pi}{2}, \varphi = \pi$) line. The radial positions of each seed are shown in panel a). From purple to dark green, streamlines starts from the front of the flux rope to its rear.}
    \label{fig:TDm_evol}
\end{figure*}

In this section, we discuss the 3D magnetic structure of TDm FRs and their evolution during their propagation of up to 210 \Rsol.
The panel a) of Figure \ref{fig:TDm_evol} shows a radial cut along the $\theta = \frac{\pi}{2}$, $\varphi = \pi$ line for the \Bph\ in red and \Bt\ in green. In panel b), the left side shows the magnetic structure of the \thicku\ FR and the right side shows the
\thint\ one. From the top to the bottom on each panel, the flux ropes are at 15, 30, 50 and finally 210 
\Rsol. Magnetic field lines start close to the front of the flux rope (purple) and goes close to its end (green). The colored dots in panel a) are the seeds of the streamlines of the corresponding color in panel b). 
We can see some magnetic field lines that reconnected with the open flux of the solar wind.

We observe that the ICMEs propagate radially. This is expected because the \B\ of the solar wind has an axisymmetry and also because the simulated ICMEs does not encounter any other ICMES. Indeed, the encounter of an ICME with another can lead to the deflection of the ICME as shown in some event (\eg \citealt{lugaz2012}) but it is not the case here.

In Figure \ref{fig:TDm_evol}, the two sides (see white arrows in Figure \ref{fig:TDm_evol}) of the \thint\ case (right of panel b)) are included in the equatorial plane, while the right (left) side of \thicku\ is above (below) the equatorial plane.
This suggests a rotation of the \thick\ case compared to the \thin\ case happening  before 15 \Rsol.  Nevertheless, we can see that the \thicku\ case seems to continue to rotate slowly after 15 \Rsol.
A similar orientation is found for FRs of the same set. It thus suggests that the initial magnetic field intensity (and thus the speed of the ICME) has no significant effect on the rotation of the FR during its propagation in our simulation.

\cite{lynch2009} studied the rotation of a flux rope formed by reconnection in numerical simulations, and they found a rotation for their left-handed flux rope (that goes in the same direction as the \thicku\ case) that reaches 50° at $3.5 \Rsol$, which is a similar angle than the one we observe at but at 15 \Rsol, see the left side of panel b) in Figure \ref{fig:TDm_evol}.

The observed rotation of the TDm flux rope is also in agreement with \cite{kay2015} who showed, with CMEs propagating in an extrapolated magnetic field with the PFSS method (potential field source surface, \citealt{schatten1969,altschuler1969}), that the rotation of the flux rope mainly occurs close to the Sun (< 10 \Rsol). Indeed, when they do not consider the torque applied on CMEs during their propagation between 10 and 210 \Rsol, they only found a 10\% error on the orientation of the CME at 210 \Rsol, implying that most of the rotation is due to the torque applied before 10 \Rsol.
Moreover, \cite{isavnin2014} studied the evolution of 14 CMEs in the heliosphere with data-driven simulations, and showed that $\approx 57 \%$ of the rotation happens below 30 \Rsol.
We essentially find the same result on the \thicku\ case, which rotates quickly before 15 \Rsol\ and then rotates slowly up to 210 \Rsol.

The observed rotation of the FR could be caused by a kink instability, as discussed by \cite{manchester2017}. However, 
the thinnest FRs, the \thin\ ones, do not seem to rotate as much as the \thick\
and the kink instability is known to be predominant for thinner flux ropes \citep{titov2014}.

On the other hand, \cite{shiota2010} (along with \citealt{cohen2010}) suggested that the reconnection with the ambient magnetic field (in particular inside the helmet streamer) is causing a rotation of the flux rope. This may be the process at heart here, since we can see in Figure \ref{fig:TDm_erup} that the flux rope partly reconnects with the magnetic field within the streamer. This could explain why we don't see a lot of rotation happening after 15 \Rsol\ since the tip of the streamer is roughly at 10 \Rsol.

We see on the left panel of Figure \ref{fig:TDm_evol} that the \Bph\ profile for both \thicku\ and \thint\ FRs does not change its shape from 15 \Rsol\ to 210 \Rsol. Conversely, we observe that \Bt\ goes progressively to lower values (compared to the minimum value of \Bph) during the propagation. This is clearly visible for the \thint\ FR while this effect is less pronounced for the \thicku\ one. Such accumulation of \Bt\ could be the trace of magnetic field lines at the rear of the ME that are reconnecting with each other. 
This reconnection process seems to be more predominant with the \thin\ FRs than for the \thick\ ones. 

The magnetic structure of the FR from the same set (\thin\ or \thick) is really similar but it seems that the two different sets show different magnetic structure.
If we compare the magnetic field lines between the \thicku\ case and the \thint\ FR in the right panel of Figure \ref{fig:TDm_evol}, they don't seem very similar. However, if we focus on the part of the streamlines that is close to the ($\theta = \frac{\pi}{2}, \varphi = \pi$) line, we observe that streamlines close to the leading edge of the FR (in purple) are roughly vertical, while the ones close to the center of the FR (in white, and light orange/green) are more horizontal. This matches the model described in \cite{lundquist1951} for a FR which axis is aligned with the ecliptic plane.

This can be seen more easily at 210 \Rsol\ but it is present at each distance presented in Figure \ref{fig:TDm_evol}. Nevertheless, streamlines show significant differences between the 2 FRs when considering the sides of the FR (even if we rotate the FR).

\section{Comparisons between PLUTO-simulated and \is\ ICMEs}
\label{sec:obs}

\subsection{Evolution of plasma and magnetic properties of the flux rope with the distance}
\label{sec:evol_param}

To study the flux rope properties, we need to identify the position of the flux rope. To do so, we developed an algorithm that fits a Gaussian on \Bph\ close to the nose of the simulated ICME at each time step. 

According to the typical \cite{lundquist1951} model, we expect the toroidal (\Bph) component to be higher than the poloidal (\Br\ and \Bt) component at the center of the flux rope. Even though the shape of \Bph\, is not necessarily expected to be  Gaussian, it is still a good approximation of the location of the flux rope. To enhance the automatic detection, we do a 2 steps Gaussian fitting. After the first fit, we find the position of the local minimum of \Bph\ around the position of the fitted Gaussian. We then use this position as an initial guess for the 2nd Gaussian fit. Finally, the position of the FR corresponds to the position of the 2nd fitted Gaussian. This fit is illustrated in Figure \ref{fig:TDm_evol} on the left panel, third row.
In order to help the fitting algorithm to find the position of the FR at the next time step, the speed at the center of the first fitted Gaussian is used to predict the next position of the ICME and thus facilitates the automated detection.

\begin{figure}
    \centering
\includegraphics[width=\linewidth]{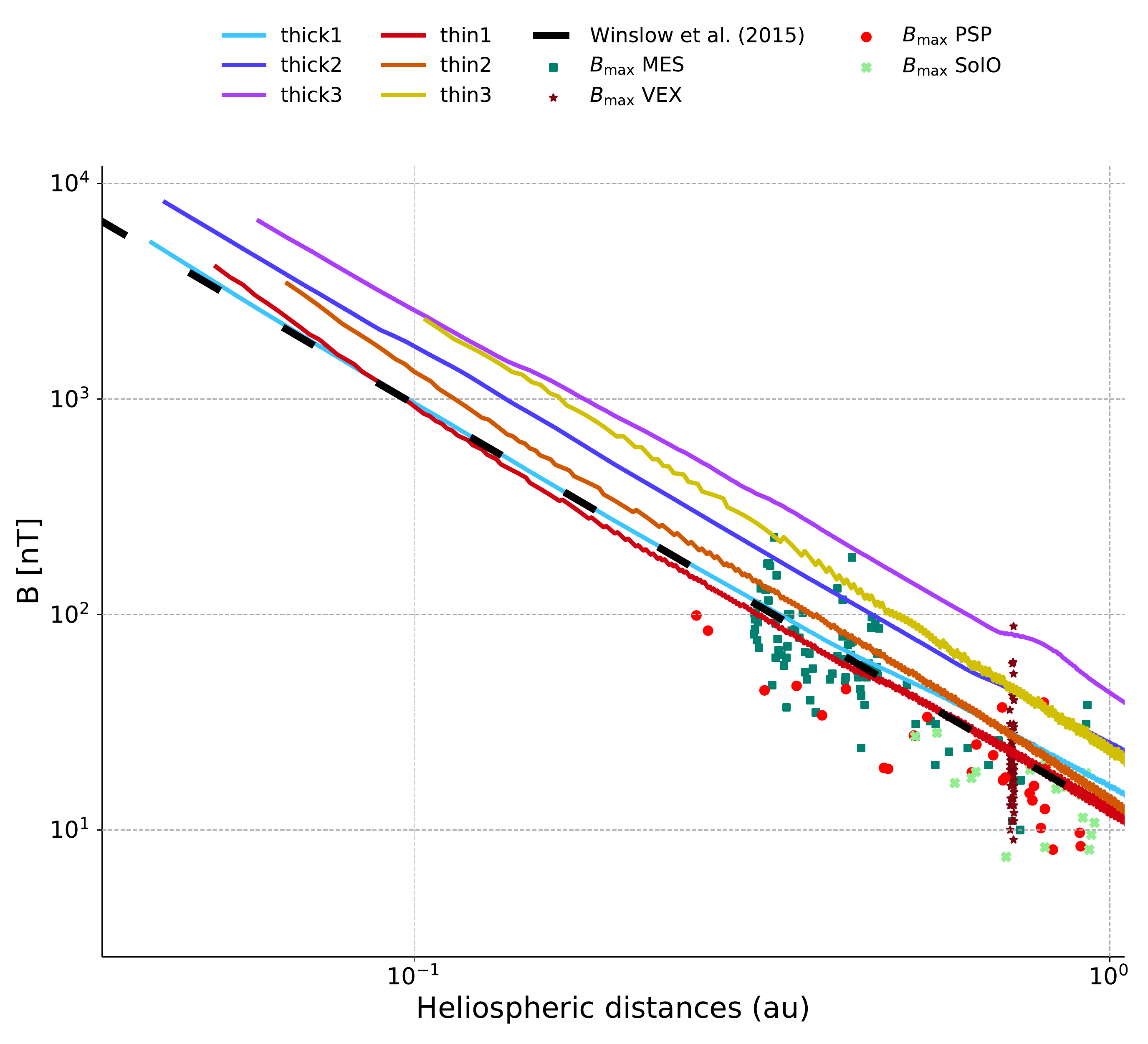}
    \caption{Evolution of the magnetic field (in nT) at the position determined by the Gaussian fitting of the ME as a function of the distance from the Sun (in au). The solid colored lines correspond to the simulated flux ropes. The black dashed line corresponds to a power law deduced from the observations in \cite{winslow2015}. Finally, the maximum magnetic field measured during an ICME encounter at Messenger (MES), Venus Express (VEX), Parker Solar Probe (PSP) and Solar Orbiter (SolO) \citep{winslow2015,good2016,mostl2017,mostl2020} are also shown in this plot.}
    \label{fig:comp_winslow}
\end{figure}

Figure \ref{fig:comp_winslow} shows the evolution of the magnetic field amplitude
measured at the position determined by the Gaussian fitting for each time step as a function of the propagation distance in log-log scale.  The
values for the six simulated flux ropes are represented by the solid colored lines.  We see that these intensities decrease like a power law along the flux rope propagation. 

Such laws have been derived in the literature to
determine the magnetic field of the flux rope at different distances from the
Sun (\eg \citealt{bothmer1997,liu2005,wang2005,leitner2007,winslow2015} and references therein). 
The
black dashed line corresponds to a power-law fit of the maximum magnetic field $B_{\rm max}$ measured during an ICME encounter
performed by \citet{winslow2015}. The authors
found a power law with an exponent of $\alpha_B = -1.89 \pm 0.14$ for the evolution of $B_{\rm max}$ as a function
of the distance with the 95\% confidence interval. We see that the
six simulated flux ropes follow roughly a similar trend. Finally, the symbols correspond to the maximum magnetic field
measured by Messenger, Venus Express, Parker Solar Probe and Solar Orbiter during an ICME encounter using the \cite{winslow2015,good2016} and the HELIO4CAST ICMECAT v11\footnote{\url{https://helioforecast.space/icmecat}}\footnote{\url{https://doi.org/10.6084/m9.figshare.6356420}} catalog \citep{mostl2017,mostl2020}.
\
We point out that the \thicku\ FR rotates according to the 3D visualization displayed in Figure \ref{fig:TDm_evol}. This might introduce a bias in the automated detection of the position of the flux rope (and thus the measurement of \B). Nevertheless, this FR is found to rotate slowly after 15 \Rsol\ meaning that the magnitude of B might be affected by this rotation but not its trend after 15 \Rsol\ as a function of distance because the \Bph\ profiles are very similar at different distance.

In order to compare the evolution of the TDm flux rope with the observations more
quantitatively, we perform a power law fit of the evolution of the simulated $B$ as a
function of distance. The corresponding slope exponents ($\alpha$) are reported in table
\ref{tab:slope_B}.

\begin{table}[h]
\begin{center}
\begin{tabular}{c c c c c c c }
    \hline
    cas & \thinu\ & \thind\ & \thint\ & \thicku\ & \thickd\ & \thickt\  \\
    \hline
     $\alpha$  & -1.860 & -2.019 & -2.064 & -1.759 & -1.828 & -1.798   \\
     $\Delta \alpha$  & 0.002 & 0.002 & 0.003 & 0.002 & 0.003 & 0.004   \\
    \hline
\end{tabular}
\caption{Slopes of the linear regression and associated error bars of the evolution law of the
    magnetic field for the 6 cases of this study. We recall that \citet{winslow2015} found $\alpha = -1.89 \pm 0.14$ based on maximum \B\ measurements during ICME encounters with \is\ data.}
\label{tab:slope_B}
\end{center}
\end{table}

We see that the slopes of all six TDm flux ropes fit within the 95\% confidence interval of the
\citet{winslow2015} power law. There is thus a good agreement between the power
law deduced from the observations and the magnetic field evolution of TDm flux
ropes.

We also find that the largest magnetic field strengths in our sample, the \thickt\ (purple line), lies on the upper range observed for ICMEs. It means that the magnetic field of the simulated ICMEs have a realistic, if somewhat high, magnetic field strength at different distances from the Sun between Mercury and the Earth.

To summarize, we find that the eruption of an initially out of equilibrium flux rope and
its propagation in a quasi-isothermal solar wind produces magnetic field amplitudes that agree with the \is\ observations of ICMEs at the different distances from the Sun.


Proceeding similarly, we compare the evolution of the magnetic field, the speed, the density, the temperature and the $\beta$ parameter with the distance with \is\ measurements made during ICME encounters from \cite{liu2005} but also with the result of \cite{scolini2021} that used EUHFORIA simulations with a spheromak magnetic structure \citep{verbeke2019}. Table \ref{tab:slope_plasma} shows the fit parameter for
the ICMEs we simulate (averaged over all the FRs) with the associated error on the row just below. The next row shows fitting parameters from \cite{scolini2021}. Finally, the 2 last rows correspond to fitting parameters and the associated error from \cite{liu2005}.

\cite{liu2005} studied ICMEs seen by different probes and performed a power law fit on their magnetic and plasma 
parameters. We find in Table \ref{tab:slope_plasma} a good agreement between \cite{liu2005} and these simulation results for the density and the temperature. The speed of the simulated ICMEs decreases faster than in the observations, but these values are still close. This difference can be explained for instance by the interaction of the FR with the solar wind. In particular, in our model the wind is a bit denser than the solar wind, which makes the propagation of the flux rope more difficult.

There is also a good match between the evolution of \B, \Vp\ and \np\ with the distance in \cite{scolini2021} and in the study presented in this paper (see third row of Table \ref{tab:slope_plasma}).
Despite the differences discussed earlier  between the TDm + PLUTO and the Linear Force Free Spheromak model (LFSS) + EUHFORIA model (see Section \ref{sec:intro}), power laws are found to be similar in both models. This shows the robustness of the result.

However, if we compare the evolution of $T$ and \be\ we find some differences.
Indeed, we find that \aT\ and \ab\ are -0.37 and 1.04 respectively in the PLUTO
simulations while their values are -1.19 and 0.11 in \euh\ simulations.
These differences are probably due to the solar wind model, which is not the same in the
PLUTO and EUHFORIA simulations. Indeed, as described in Section \ref{sec:solarwind}
the solar wind model used in this study is quasi-isothermal. It means that
the temperature is almost constant in the whole simulation domain of PLUTO. This
explains why the temperature evolution is different compared to the \euh\ simulations,
which have a more realistic description of the thermal expansion of the solar wind.

\begin{table}[t!]
\begin{center}
\resizebox{\linewidth}{!}{
\begin{tabular}{l c c c c c }
    \hline
     Source & \aB & \aV & \an & \aT & \ab \\
    \hline
    PLUTO + TDm & -1.88 & -0.1 & -2.37 & -0.37 & 1.04 \\
    $\Delta$  & 0.05 & 0.03 & 0.05 & -0.1 & 0.06 \\ \hline
    EUHFORIA + LFSS & -1.90 & -0.08 & -2.38 & -1.19 & 0.11 \\ \hline
    \citet{liu2005} & -1.40 & 0.002 &  -2.32 & -0.36 & - \\
    $\Delta$  & 0.08 & 0.02 &  0.07 & 0.06 & -  \\
    \hline
\end{tabular}
}
\caption{Power-law exponent for the evolution
of the magnetic field, the speed, the density, the temperature and the \be\ (when applicable) with
the distance from the Sun and their errors averaged over the 6 TDm flux ropes, from the \cite{scolini2021} study (third row) and from the \cite{liu2005} and the associated error (fourth and last row).}
\label{tab:slope_plasma}
\end{center}
\end{table}

\subsection{Synthetic crossing}
\label{sec:SC}

We now compare the properties of the propagation of a TDm flux rope in a
quasi-isothermal solar wind with \is\ profiles made by spacecraft during an
ICME encounter. To do so, synthetic crossings are performed through the flux rope during its propagation, and are compared with the observations performed at 1 au by ACE in \cite{regnault2020}.
This study used the superposed epoch method on more than
300 ICMEs and computed the most probable profile for different physical parameters, which can then be used as a typical crossing profile to assess the realism of our modeled ICMEs.

\begin{figure*}
    \centering
    \includegraphics[width=0.8\linewidth]{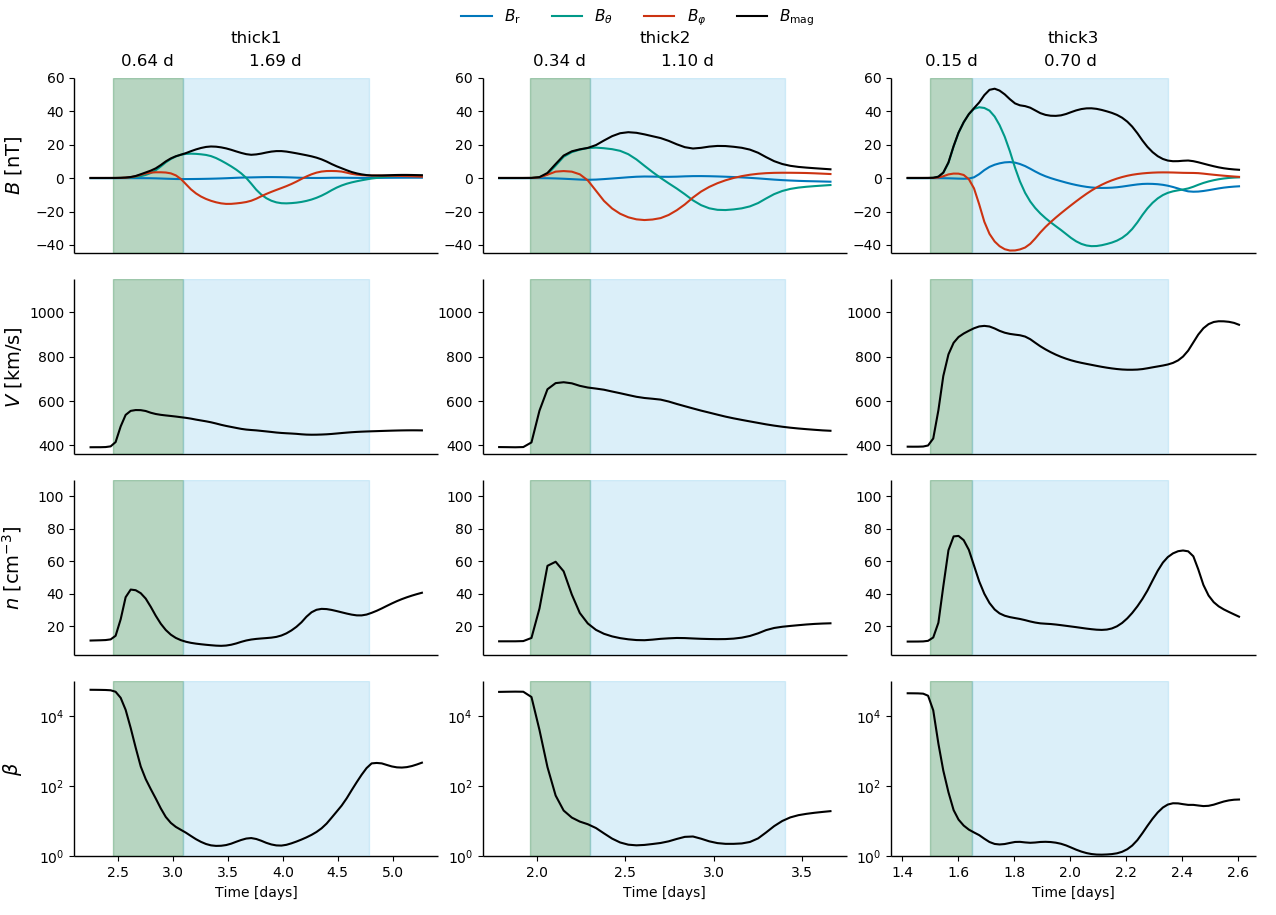}
    \caption{Synthetic crossings at 1 au of the three \thick\ flux ropes. From the top to bottom we show the magnetic field and its components (see the legend for the color code) in nT, the speed in km/s, the density in \cm and the \be\ parameter as a function of time. The green and the blue areas correspond respectively to the sheath and the ME.}
    \label{fig:SC_thick}
\end{figure*}

\begin{figure*}
    \centering
    \includegraphics[width=0.8\linewidth]{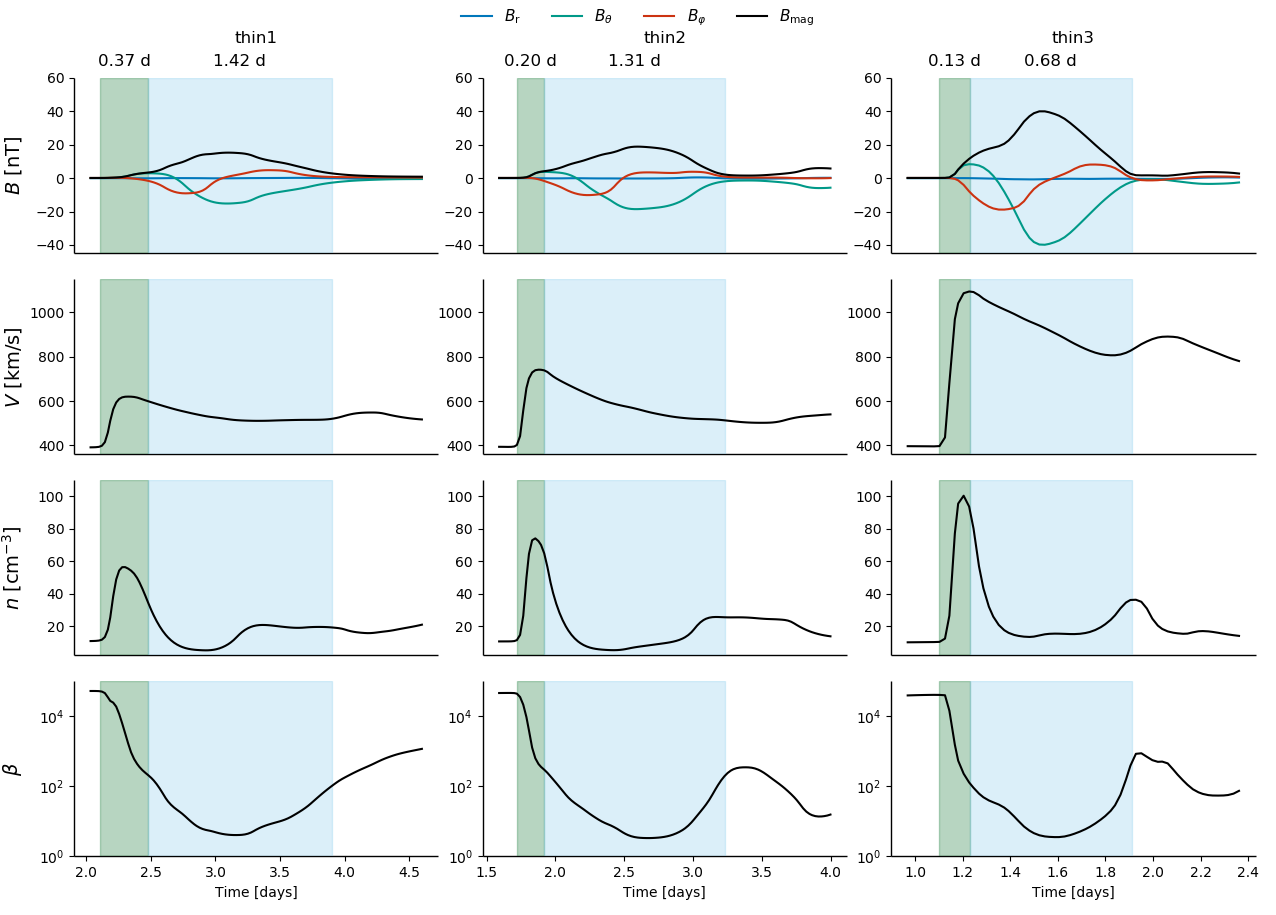}
    \caption{Synthetic crossings at 1 au of the three \thin\ flux ropes. The physical parameter and the color code are the same as in Figure \ref{fig:SC_thick}. }
    \label{fig:SC_thin}
\end{figure*}

Figures \ref{fig:SC_thick} and \ref{fig:SC_thin} show synthetic crossings at 210
\Rsol\ at ($\varphi = \pi$, $\theta = \frac{\pi}{2}$). The
green and blue areas show the sheath and magnetic ejecta (ME) areas respectively, as they would have been
defined in actual \is\ data according to the typical signatures detailed in Section \ref{sec:intro}. 
From top to bottom, we show the magnetic field and
its components in nT, the speed in km/s, the density in \cm\ and the plasma
\be. We observe in these two figures the typical expected ICME signatures, which we now detail.

The beginning of the sheath, a region of compressed and heated plasma, is
highlighted by a sudden increase of the speed and the density. This
increase is also present for the magnetic field, but it is smoother. We note that the increase of the speed
and density are not as sudden as we can observe in \is\ measurements where there is a
discontinuity in the data at the beginning of the sheath (\eg see
ICME examples in \citealt{masias-meza2016}). The fact that the increase of \V\ and $n$ is smoothed
might come from the HLL Riemann solver that we chose, which is quite diffusive.
Moreover, it might also be caused by the spatial resolution which can cause numerical
diffusion, especially that far from the Sun where the grid is coarser. As mentioned earlier, the finer spatial resolution at 1 au is $\env 2.5$ \Rsol.

More specifically, the density in
the sheath is increased compared to the value of the solar wind by a factor $\approx 3$ for \thicku, 4
for \thickd, 5.3 for \thickt, 4 for \thinu, 5 for
\thind\ and finally 6.7 for \thint.
\cite{janvier2014a} studied the change of properties of the solar wind after
shocks detected in \is\ data. In particular, they found that ICME-driven shocks
produce an increase of density that lies between 1.4 and 4.6. 
Hence, cases \thicku, \thickd\ and \thinu\ agree well with these observations.
We also observe that the faster the ICME propagates, the denser the sheath. The solar wind accumulates more in front of the flux rope since it has less time to evacuate on the sides.

We observe that the \be\ parameter decreases after the beginning of the sheath
and stays lower than the solar wind value up to the end of the magnetic ejecta.
However, in \cite{regnault2020} the authors found that the
plasma \be\ remains almost constant in the sheath and starts to decrease only
in the ME. Moreover, we find a \be\ which is really high in the simulation compared to the measured with \is\ data. This is due to the effect of our quasi-isothermal approximation, which leads to unrealistically high thermal pressures and increases the values of \be. A 1 au, the modelled ICMEs have about 4 orders of magnitude too high of a pressure compared to
\cite{masias-meza2016,nieves-chinchilla2018,regnault2020}. 
With the \be\ parameter that stays $> 1$ inside the ME for all the FR (instead of $< 1$ as observed in \is\ profile) it means that in the simulation the plasma forces are dominant (compared to the magnetic ones). This extra plasma pressure could cause for instance an "over" expansion (compared with expansion caused by the magnetic pressure only).
These issues are the main weaknesses of this model. However, they could be alleviated using a more realistic solar wind heating process (based \eg\ on Alfv\'en-wave heating, see \citealt{reville2020}) which is out of the scope of the present paper and will be explored in future studies. 

The synthetic \B\ profile of \thick\ FRs shows a double bump structure in
the ME, one right after the beginning of the ME and a second one close to the
end. Conversely, \thin\ cases show only one bump. However, we observe the
smooth rotation of the $\theta$ component of the magnetic field in each case,
showing the signature of a magnetic flux rope. The beginning of the ME corresponds to the start of the rotation of \Bt. The fact that the rotation is smooth and almost 
symmetric (from \env 20 nT to -20 nT for the \thickd\ case for example), and the fact that \Bph\ is at its maximum (in absolute value) almost when \Bt\ = 0, matches very well with the idea of a flux rope according to \cite{lundquist1951}. This agrees with the similarities discussed in Section \ref{sec:prop} between the part of the streamlines that are close to the FR for the \thin\ and \thick\ FRs. However, this local agreement does not imply that both magnetic structures are actually similar as we can see in Figure \ref{fig:TDm_evol}.

We also observe that \Bt\ crosses 0 almost when \Bph\ is at its maximum at the very beginning of the \thin\ MEs.
However, we see that the \Bt\ rotation for the \thin\ FRs is clearly asymmetric, as pointed out in Section \ref{sec:prop}.

The beginning of the ME matches well with the location where the density goes back to similar value than that of the solar wind for the \thicku and \thickd\ cases. It is nevertheless not the case for the \thin\ MEs. This could be explained by reconnection happening between the sheath and the ME magnetic field allowing plasma to enter the FR.

We can see that the magnitude of \B\ of \thinu\ and \thicku\ is similar, $\env 20$ nT at the front part of the ME while the initial \B\ presented in Table \ref{tab:cases} shows that the \thicku\ case is roughly 2 times more magnetized than the \thinu. The same behavior is observed for the \thickd\ and \thind\ FRs. This can be explained by a greater expansion of the magnetic structure of \thin\ FRs during their propagation and thus greater decrease of the inner \B.

\cite{regnault2020} used the relative speed of the ICME compared to the solar wind to build different groups of ICME. They found that the "relatively fast" ICMEs (fast compared to the local solar wind) have an asymmetric magnetic field profile with a higher magnetic field at the front of the ME as shown in their Figure 4 . However, in our simulations, for which all the simulated 6 FRs would be considered as relatively fast, we don't find such an asymmetry. 

We observe that the speed follows a monotonically decreasing profile. This
is due to the radial expansion of the flux rope, as often observed in \is\ data (see \eg ICME examples in \citealt{farrugia1993,masias-meza2016,regnault2020}). Moreover, the density goes back to
pre-ICME solar wind values for most of the magnetic ejecta as was observed in
\cite{regnault2020}. The part of the ME that shows the smooth rotation matches well with the region of the lowest density for the \thinu\ and \thicku\ MEs. This low density probably highlights the part of the magnetic ejecta that has the shape of a FR, as suggested by the smooth rotation of \Bt. The remainder of the magnetic ejecta corresponds to the eroded part of the flux rope due to reconnection with the solar wind as described by \cite{dasso2006,ruffenach2012} in their Figure 6 and Figure 1 respectively. As pointed out by Figure \ref{fig:TDm_erup}, the initial \B\ has an impact on the velocity of the FR during its eruption. This relation between the initial \B\ and the speed is still clear at 1 au in one set of the simulated FR. 

In \cite{regnault2020}, the authors found that the mean, median and the most probable value of the relative 
duration of the ME compared with the sheath range from 1.5 to 3.5.
In our simulation, the duration of the sheath and the ME is shown in days on 
top of each column of Figures \ref{fig:SC_thick} and \ref{fig:SC_thin}.
By computing the ratio of the ME duration relatively to the sheath we find 2.6, 3.2 and 5.0 for the \thicku, 
\thickd\ and \thickt\ events respectively, while we have 3.8, 6.55 and 5.2 for the \thinu, \thind\ and the \thint\ 
events. We thus found a good match between the simulation and the observation, except for the \thickt, \thind\ and \thint\ FRs which has a higher ratio than the average.
Finally, we also observe that the sheath is shorter (compared to the ME) for faster FRs for the \thick\ set. We interpret that as an enhanced compression of the sheath due to the higher speed. This behavior is less clear for the \thin\ set if we take the duration of the sheath, using the \Bt\ rotation as a start for the ME. However, we can see that the width at half height of the density peak is lower for faster \thin\ FR. Suggesting a higher compression for faster \thin\ FRs, as for the \thick\ FRs.

\subsection{Effect of rotation on the identification of the magnetic ejecta}
\label{sec:SC_rot}

As illustrated in Figure \ref{fig:TDm_evol}, the \thicku\ FR rotates 
during its propagation, while the \thin\ FR axis remains aligned with the equator.
The relative orientation of \thick\ FRs with respect to \thin\ FRs is of importance if we want to compare the 
components of \B.

\begin{figure*}
    \centering
    \includegraphics[width=\linewidth]{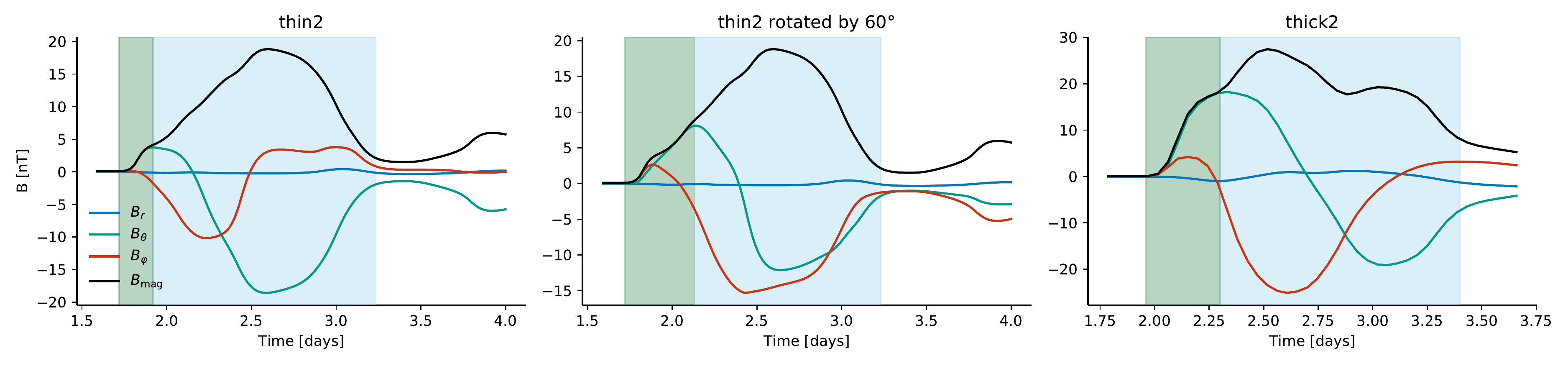}
    \caption{Synthetic crossing at 1 au of the \thind\ and the \thickd\ From the left to the right, \B and its components (same color code as in Figure \ref{fig:SC_thick}, as for the colored area) are shown for the \thind\ FR, for the \thind\ with a 60° of rotation on the components and finally for the \thickd\ case for comparison.}
    \label{fig:SC_rot}
\end{figure*}

Figure \ref{fig:SC_rot} shows the effect of the rotation on the \B\ components at 1 au.
From the left to the right, \B and its components are shown for the \thind\ FR, for the \thind\ with a 60° of rotation on the components and finally for the \thickd\ case for comparison. The angle 60° has been found simply finding the angle that maximizes the symmetry (estimated by eye) of the \Bt\ rotation looking at angles between 0 and 90° with 10° of increment. We note that $\pm 10$° does not significantly change the results that we describe just after.

The rotation of the magnetic field components of the \thin\ (illustrated here only for \thind) allows us to obtain synthetic crossings which are very similar between the \thin\ and the \thick\ FRs. In both cases (middle and right panels) the \Bt\ rotation is  symmetric and \Bph\ is at its maximum when \Bt\ is close to 0.

We also note that \Bt\ starts its rotation later. Indeed, instead of starting at $t = 1.92$ days, it starts at $t=2.13$ days. Since we use \Bt\ smooth rotation as parameter to define the ME boundaries, the rotation of the magnetic field component has an impact on their definitions. In the new coordinate system, the rotated ME has a lower density compared to its non-rotated version. Such a lower density is in better agreement with 
the typical sheath / ME transition as observed in the typical profiles \citep{regnault2020}. Finally, it also produces shorter ME over sheath duration which allows the \thin\ FRs to better match with the \cite{janvier2014a} study. 

Figures \ref{fig:TDm_evol} and \ref{fig:SC_rot} show that, in our simulations, the flux ropes that are tilted with respect to the equator 
(the \thick\ cases, according to the 3D visualization) show FR signatures that are more clear than the FR that is aligned with the equator while crossed at the nose. 
If we apply a rotation on the FRs that are aligned with the equator, we recover clearer FR signatures. 

We see here that the synthetic crossing of the \B\ components at the nose of the FR shows similar features between the \thint\ and the \thicku\ FRs if we apply a rotation to one of the FR. Nevertheless, the magnetic structure presented in Figure \ref{fig:TDm_evol} shows 
significant differences in the sides of the FR.
This highlights that the resulting profiles are very dependent on the crossing location within the structure. To lift this ambiguity, multiple crossings of the ICME are necessary.

\section{Summary and Discussion}
\label{sec:discussion}

We have presented a novel numerical model that simulates the propagation of a TDm flux rope in a quasi-isothermal solar wind from the low corona to 2 au. We have applied this model to a series of TDm considering for now a solar wind with a dipolar magnetic field configuration resembling solar minimum phases. The flux ropes are initialized out of equilibrium such that they erupt quickly after their insertion.

During their propagation, we compare the evolution of physical properties of the ICMEs (such as their magnetic field, speed and 
density) with evolution laws of these parameters deduced from \is\ observations. The evolution of the magnetic field of the 
simulated ICMEs follows the same trend as the evolution of $B_{\rm max}$  measured by space 
probes during an ICME encounter at different distances \citep{winslow2015}. Also, if we compare directly the value of the magnetic field in the 
simulations with $B_{\rm max}$ measured by different probes (such as Messenger, Venus Express, Solar Orbiter and Parker Solar 
probe) we find that the magnetic field of the simulated ICMEs is comparable to actual measurements in ICMEs, with the \thickt\ flux rope that lies on the 
upper edge of the range of observations. We also find a good agreement on the evolution of the density and the temperature with the study of \citet{liu2005}.

We also find a good match between the evolution of \B, \Vp, \np\ with the distance between the PLUTO + TDm simulation and the EUHFORIA + spheromak simulation. This shows that even if the initial magnetic structures are different between the two simulations and are initialized at different distances, the evolution of \B\ in the heliosphere is similar.

However, we do find differences in the temperature and the \be\ parameter with are due to the quasi-isothermal assumption done in this study and not in the simulations of \cite{scolini2021}.

The change of the initial thickness of the TDm flux rope has a significant impact on the properties of the flux rope while it propagates. The first main effect is that flux ropes with different initial sizes are subject to different rotations. Thicker flux ropes rotate more in this study, which suggests that the kink instability is not the main process at the heart of the observed rotation. We interpret the rotation as a consequence of reconnection happening between the magnetic structure and the ambient solar wind. The main part of the rotation takes place in the low corona (<15 \Rsol).

We also compare synthetic crossings done close to the nose of the simulated ICMEs with the superposed-epoch analysis of
\cite{regnault2020}. Overall, the typically expected ICME signatures are observed in the simulation. We observe an increase of 
the magnetic field, the speed and the density in the sheath area (with a smooth increase for \B) due to the compression of the 
solar wind plasma that accumulates in front of the propagating flux rope. We however find that the increase of \B\ at the beginning of the sheath is smoother than for \np\ and \Vp.
In the magnetic ejecta we find a lower of density compared to the 
sheath area, and a monotonic decrease of the speed which are interpreted as signs of the expansion of the simulated flux rope during its propagation. We observe a smooth rotation of \Bt\ which is the trace of the TDm flux rope that is still observable at 1 au. The \be\ parameter starts to decrease at the beginning of the sheath, while the \be\ of the sheath of the superposed epoch profile in \cite{regnault2020} remains the same as the solar wind.
We also see that its value is higher than the one measured at 1 au. This is due to the quasi-isothermal assumption  that produces a strong thermal pressure over the whole simulation domain and thus increasing the \be\ parameter.
This highlights the need for a better solar wind model, like the WindPredict-AW model \citep{reville2020} that has a more realistic thermal treatment of the solar corona.

The two sets of flux rope \thin\ and \thick\ do not show any clear difference in their 3D magnetic structure close to their noses, except for their rotation during their propagation. Similarly, the synthetic crossings do not show any clear difference between the 2 sets. Even the components of the magnetic field (once the rotation is corrected) are fairly similar. There is thus no clear effect of the initial size of the flux rope on the observational properties at 1 au close to the nose of the FR. Indeed, the duration of the modelled events is similar at 1 au disregarding their initial size. However, differences are expected between the 2 set of FR when far from the nose, highlighting the effect of the crossing on ICME properties.

The initial magnetic field does not impact significantly the rotation of the flux rope during its propagation, as FR sharing the same size exhibit the same rotation pattern.
Moreover, the change of initial magnetic field have an impact on the speed of the simulated ICMEs due to greater Lorentz forces happening
during the eruption of the flux rope. It also leads to a higher compression of the solar wind plasma and thus a higher density in the sheath area.

To conclude, the eruption and propagation of an initially out of equilibrium TDm flux rope manages to produce a magnetic structure that has the typical signatures as seen in \is\ data.
The results obtained during this study highlight the difficulty to determine the 3D magnetic structure of an ICME because of the degeneracy of unique \is\ profiles for each ICME. It thus stresses out the need for multispacecraft observations of ICMEs in order to determine their global properties. 
Thanks to the launch of Parker Solar Probe \citep{fox2016},  Solar Orbiter \citep{muller2020}, and Bepi Colombo \citep{benkhoff2021} we have now more chances to characterize ICMEs propagation with multipoint in-situ observations as described in \citep{hadid2021}. One example of such study using Solar Orbiter, Bepi Colombo and Wind was for instance carried out in \citet{davies2021a}.

As a follow-up of this study, more numerical work will be done. First, to insert the TDm FR in a more realistic solar wind and secondly, to perform the relaxation of 
the flux rope before its eruption is triggered. This could allow us to study in more details 
the physical mechanism that lead to the eruption of such a flux rope in a full 3D MHD model, 
as well as the initial rotation in the low corona when the erupting structure interacts with 
the ambient magnetic field and how its topology impact the property of the propagating FR.

\begin{acknowledgements}
F. R., M. J., A. S. and F. A. acknowledge
financial support from the Dim-ACAV
Île-de-France region doctoral grant as
well as support from the EDOM
Doctoral School at Université
Paris-Saclay. Computations were carried out using CEA TGCC and CNRS IDRIS facilities within the GENCI 80410133 and 100410133 allocations, and a local meso-computer founded by DIM ACAV+. A. S. acknowledges funding from the Programme National Soleil-Terre (PNST).
F.R., N. L. N. A. acknowledge the financial support of 80NSSC21K0463 and 80NSSC20K0700 funding.
\end{acknowledgements}

\bibliography{Research}

\begin{thebibliography}{82}
\expandafter\ifx\csname natexlab\endcsname\relax\def\natexlab#1{#1}\fi

\bibitem[{{Al-Haddad} {et~al.}(2019){Al-Haddad}, Poedts, Roussev, Farrugia, Yu,
  \& Lugaz}]{al-haddad2019a}
{Al-Haddad}, N., Poedts, S., Roussev, I., {et~al.} 2019, ApJ, 870, 100

\bibitem[{Altschuler \& Newkirk(1969)}]{altschuler1969}
Altschuler, M.~D. \& Newkirk, G. 1969, Sol Phys, 9, 131

\bibitem[{Aulanier(2010)}]{aulanier2010}
Aulanier, G. 2010, 708, 20

\bibitem[{Aulanier {et~al.}(2012)Aulanier, Janvier, \&
  Schmieder}]{aulanier2012}
Aulanier, G., Janvier, M., \& Schmieder, B. 2012, A\&A, 543, A110

\bibitem[{Benkhoff {et~al.}(2021)Benkhoff, Murakami, Baumjohann, Besse, Bunce,
  Casale, Cremosese, Glassmeier, Hayakawa, Heyner, Hiesinger, Huovelin,
  Hussmann, Iafolla, Iess, Kasaba, Kobayashi, Milillo, Mitrofanov, Montagnon,
  Novara, Orsini, Quemerais, Reininghaus, Saito, Santoli, Stramaccioni,
  Sutherland, Thomas, Yoshikawa, \& Zender}]{benkhoff2021}
Benkhoff, J., Murakami, G., Baumjohann, W., {et~al.} 2021, Space Sci Rev, 217,
  90

\bibitem[{Bothmer \& Schwenn(1997)}]{bothmer1997}
Bothmer, V. \& Schwenn, R. 1997, 24

\bibitem[{Burlaga {et~al.}(1981)Burlaga, Sittler, Mariani, \&
  Schwenn}]{burlaga1981}
Burlaga, L., Sittler, E., Mariani, F., \& Schwenn, R. 1981, Journal of
  Geophysical Research: Space Physics, 86, 6673

\bibitem[{Carmichael(1964)}]{carmichael1964}
Carmichael, H. 1964, NASA Special Publication, 50

\bibitem[{Casini {et~al.}(2003)Casini, Ariste, Tomczyk, \& Lites}]{casini2003}
Casini, R., Ariste, A.~L., Tomczyk, S., \& Lites, B.~W. 2003, ApJ, 598, L67

\bibitem[{Chen(2011)}]{chen2011}
Chen, P.~F. 2011, Living Reviews in Solar Physics, 8

\bibitem[{Chiu {et~al.}(1998)Chiu, {Von-Mehlem}, Willey, Betenbaugh, Maynard,
  Krein, Conde, Gray, Hunt, Mosher, McCullough, Panneton, Staiger, \&
  Rodberg}]{chiu1998}
Chiu, M., {Von-Mehlem}, U., Willey, C., {et~al.} 1998, Space Science Reviews,
  86, 257

\bibitem[{Cohen {et~al.}(2010)Cohen, Attrill, Schwadron, Crooker, Owens, Downs,
  \& Gombosi}]{cohen2010}
Cohen, O., Attrill, G. D.~R., Schwadron, N.~A., {et~al.} 2010, Journal of
  Geophysical Research: Space Physics, 115

\bibitem[{Dasso {et~al.}(2006)Dasso, Mandrini, D{\'e}moulin, \&
  Luoni}]{dasso2006}
Dasso, S., Mandrini, C.~H., D{\'e}moulin, P., \& Luoni, M.~L. 2006, A\&A, 455,
  349

\bibitem[{Davies {et~al.}(2021)Davies, M{\"o}stl, Owens, Weiss, Amerstorfer,
  Hinterreiter, Bauer, Bailey, Reiss, Forsyth, Horbury, O'Brien, Evans,
  Angelini, Heyner, Richter, Auster, Magnes, Baumjohann, Fischer, Barnes,
  Davies, \& Harrison}]{davies2021a}
Davies, E.~E., M{\"o}stl, C., Owens, M.~J., {et~al.} 2021, A\&A, 656, A2

\bibitem[{D{\'e}moulin(2009)}]{demoulin2009}
D{\'e}moulin, P. 2009, Solar Physics, 257, 169

\bibitem[{D{\'e}moulin {et~al.}(2013)D{\'e}moulin, Dasso, \&
  Janvier}]{demoulin2013}
D{\'e}moulin, P., Dasso, S., \& Janvier, M. 2013, Astronomy \& Astrophysics,
  550, A3

\bibitem[{Farrugia {et~al.}(1993)Farrugia, Burlaga, Osherovich, Richardson,
  Freeman, Lepping, \& Lazarus}]{farrugia1993}
Farrugia, C.~J., Burlaga, L.~F., Osherovich, V.~A., {et~al.} 1993, Journal of
  Geophysical Research: Space Physics, 98, 7621

\bibitem[{Finley {et~al.}(2018)Finley, Matt, \& See}]{finley2018}
Finley, A.~J., Matt, S.~P., \& See, V. 2018, ApJ, 864, 125

\bibitem[{Fox {et~al.}(2016)Fox, Velli, Bale, Decker, Driesman, Howard, Kasper,
  Kinnison, Kusterer, Lario, Lockwood, McComas, Raouafi, \& Szabo}]{fox2016}
Fox, N.~J., Velli, M.~C., Bale, S.~D., {et~al.} 2016, Space Sci Rev, 204, 7

\bibitem[{Gold \& Hoyle(1960)}]{gold1960}
Gold, T. \& Hoyle, F. 1960, Monthly Notices of the Royal Astronomical Society,
  120, 89

\bibitem[{Good \& Forsyth(2016)}]{good2016}
Good, S.~W. \& Forsyth, R.~J. 2016, Sol Phys, 291, 239

\bibitem[{Gulisano {et~al.}(2010)Gulisano, D{\'e}moulin, Dasso, Ruiz, \&
  Marsch}]{gulisano2010}
Gulisano, A.~M., D{\'e}moulin, P., Dasso, S., Ruiz, M.~E., \& Marsch, E. 2010,
  A\&A, 509, A39

\bibitem[{Hadid {et~al.}(2021)Hadid, G{\'e}not, Aizawa, Milillo, Zender,
  Murakami, Benkhoff, Zouganelis, Alberti, Andr{\'e}, Bebesi, Califano,
  Dimmock, Dosa, Escoubet, Griton, Ho, Horbury, Iwai, Janvier, Kilpua, Lavraud,
  Madar, Miyoshi, M{\"u}ller, Pinto, Rouillard, Raines, Raouafi, Sahraoui,
  {S{\'a}nchez-Cano}, Shiota, Vainio, \& Walsh}]{hadid2021}
Hadid, L.~Z., G{\'e}not, V., Aizawa, S., {et~al.} 2021, Front. Astron. Space
  Sci., 8, 718024

\bibitem[{Harten \& Clark(1995)}]{harten1995}
Harten, R. \& Clark, K. 1995, Space Sci Rev, 71, 23

\bibitem[{Hirayama(1974)}]{hirayama1974}
Hirayama, T. 1974, Sol Phys, 34, 323

\bibitem[{Isavnin {et~al.}(2014)Isavnin, Vourlidas, \& Kilpua}]{isavnin2014}
Isavnin, A., Vourlidas, A., \& Kilpua, E. K.~J. 2014, Sol Phys, 289, 2141

\bibitem[{Janvier(2017)}]{janvier2017}
Janvier, M. 2017, J. Plasma Phys., 83, 535830101

\bibitem[{Janvier {et~al.}(2014)Janvier, D{\'e}moulin, \& Dasso}]{janvier2014a}
Janvier, M., D{\'e}moulin, P., \& Dasso, S. 2014, Astronomy \& Astrophysics,
  565, A99

\bibitem[{Janvier {et~al.}(2019)Janvier, Winslow, Good, Bonhomme, D{\'e}moulin,
  Dasso, M{\"o}stl, Lugaz, Amerstorfer, Soubri{\'e}, \& Boakes}]{janvier2019}
Janvier, M., Winslow, R.~M., Good, S., {et~al.} 2019, Journal of Geophysical
  Research: Space Physics, 124, 812

\bibitem[{Jian {et~al.}(2006)Jian, Russell, Luhmann, \& Skoug}]{jian2006}
Jian, L., Russell, C.~T., Luhmann, J.~G., \& Skoug, R.~M. 2006, Sol Phys, 239,
  393

\bibitem[{Kaiser \& Adams(2007)}]{kaiser2007}
Kaiser, M.~L. \& Adams, W.~J. 2007, in 2007 {{IEEE Aerospace Conference}}, 1--8

\bibitem[{Kay \& Opher(2015)}]{kay2015}
Kay, C. \& Opher, M. 2015, ApJ, 811, L36

\bibitem[{Kilpua {et~al.}(2011)Kilpua, Jian, Li, Luhmann, \&
  Russell}]{kilpua2011}
Kilpua, E., Jian, L., Li, Y., Luhmann, J., \& Russell, C. 2011, Journal of
  Atmospheric and Solar-Terrestrial Physics, 73, 1228

\bibitem[{Kilpua {et~al.}(2020)Kilpua, Fontaine, Good, {Ala-Lahti}, Osmane,
  Palmerio, Yordanova, Moissard, Hadid, \& Janvier}]{kilpua2020}
Kilpua, E. K.~J., Fontaine, D., Good, S.~W., {et~al.} 2020, Ann. Geophys., 38,
  999

\bibitem[{Kilpua {et~al.}(2015)Kilpua, Lumme, Andreeova, Isavnin, \&
  Koskinen}]{kilpua2015}
Kilpua, E. K.~J., Lumme, E., Andreeova, K., Isavnin, A., \& Koskinen, H. E.~J.
  2015, Journal of Geophysical Research: Space Physics, 120, 4112

\bibitem[{Klein \& Burlaga(1982)}]{klein1982}
Klein, L.~W. \& Burlaga, L.~F. 1982, Journal of Geophysical Research: Space
  Physics, 87, 613

\bibitem[{Kliem {et~al.}(2012)Kliem, T{\"o}r{\"o}k, \& Thompson}]{kliem2012}
Kliem, B., T{\"o}r{\"o}k, T., \& Thompson, W.~T. 2012, Sol Phys, 281, 137

\bibitem[{Kopp \& Pneuman(1976)}]{kopp1976}
Kopp, R. \& Pneuman, G. 1976, Sol Phys, 50

\bibitem[{Lanabere {et~al.}(2020)Lanabere, Dasso, D{\'e}moulin, Janvier,
  Rodriguez, \& {Mas{\'i}as-Meza}}]{lanabere2020}
Lanabere, V., Dasso, S., D{\'e}moulin, P., {et~al.} 2020, A\&A, 635, A85

\bibitem[{Leitner {et~al.}(2007)Leitner, Farrugia, M{\"o}stl, Ogilvie, Galvin,
  Schwenn, \& Biernat}]{leitner2007}
Leitner, M., Farrugia, C.~J., M{\"o}stl, C., {et~al.} 2007, Journal of
  Geophysical Research: Space Physics, 112

\bibitem[{Liu {et~al.}(2005)Liu, Richardson, \& Belcher}]{liu2005}
Liu, Y., Richardson, J., \& Belcher, J. 2005, Planetary and Space Science, 53,
  3

\bibitem[{Lugaz {et~al.}(2012)Lugaz, Farrugia, Davies, M{\"o}stl, Davis,
  Roussev, \& Temmer}]{lugaz2012}
Lugaz, N., Farrugia, C.~J., Davies, J.~A., {et~al.} 2012, ApJ, 759, 68

\bibitem[{Lundquist(1951)}]{lundquist1951}
Lundquist, S. 1951, Physical Review, 83, 307

\bibitem[{Lynch {et~al.}(2009)Lynch, Antiochos, Li, Luhmann, \&
  DeVore}]{lynch2009}
Lynch, B.~J., Antiochos, S.~K., Li, Y., Luhmann, J.~G., \& DeVore, C.~R. 2009,
  ApJ, 697, 1918

\bibitem[{Manchester {et~al.}(2017)Manchester, Kilpua, Liu, Lugaz, Riley,
  T{\"o}r{\"o}k, \& Vr{\v s}nak}]{manchester2017}
Manchester, W., Kilpua, E. K.~J., Liu, Y.~D., {et~al.} 2017, Space Sci Rev,
  212, 1159

\bibitem[{{Mas{\'i}as-Meza} {et~al.}(2016){Mas{\'i}as-Meza}, Dasso,
  D{\'e}moulin, Rodriguez, \& Janvier}]{masias-meza2016}
{Mas{\'i}as-Meza}, J.~J., Dasso, S., D{\'e}moulin, P., Rodriguez, L., \&
  Janvier, M. 2016, Astronomy \& Astrophysics, 592, A118

\bibitem[{Mignone {et~al.}(2012)Mignone, Zanni, Tzeferacos, {van Straalen},
  Colella, \& Bodo}]{mignone2012}
Mignone, A., Zanni, C., Tzeferacos, P., {et~al.} 2012, The Astrophysical
  Journal Supplement Series, 198, 7

\bibitem[{Moissard {et~al.}(2019)Moissard, Fontaine, \& Savoini}]{moissard2019}
Moissard, C., Fontaine, D., \& Savoini, P. 2019, Journal of Geophysical
  Research: Space Physics, 124, 8208

\bibitem[{M{\"o}stl {et~al.}(2017)M{\"o}stl, Isavnin, Boakes, Kilpua, Davies,
  Harrison, Barnes, Krupar, Eastwood, Good, Forsyth, Bothmer, Reiss,
  Amerstorfer, Winslow, Anderson, Philpott, Rodriguez, Rouillard, Gallagher,
  {Nieves-Chinchilla}, \& Zhang}]{mostl2017}
M{\"o}stl, C., Isavnin, A., Boakes, P.~D., {et~al.} 2017, Space Weather, 15,
  955

\bibitem[{M{\"o}stl {et~al.}(2020)M{\"o}stl, Weiss, Bailey, Reiss, Amerstorfer,
  Hinterreiter, Bauer, McIntosh, Lugaz, \& Stansby}]{mostl2020}
M{\"o}stl, C., Weiss, A.~J., Bailey, R.~L., {et~al.} 2020, ApJ, 903, 92

\bibitem[{M{\"u}ller {et~al.}(2020)M{\"u}ller, St.~Cyr, Zouganelis, Gilbert,
  Marsden, {Nieves-Chinchilla}, Antonucci, Auch{\`e}re, Berghmans, Horbury,
  Howard, Krucker, Maksimovic, Owen, Rochus, {Rodriguez-Pacheco}, Romoli,
  Solanki, Bruno, Carlsson, Fludra, Harra, Hassler, Livi, Louarn, Peter,
  Sch{\"u}hle, Teriaca, {del Toro Iniesta}, {Wimmer-Schweingruber}, Marsch,
  Velli, De~Groof, Walsh, \& Williams}]{muller2020}
M{\"u}ller, D., St.~Cyr, O.~C., Zouganelis, I., {et~al.} 2020, A\&A, 642, A1

\bibitem[{{Nieves-Chinchilla} {et~al.}(2018){Nieves-Chinchilla}, Vourlidas,
  Raymond, Linton, {Al-haddad}, Savani, Szabo, \&
  Hidalgo}]{nieves-chinchilla2018}
{Nieves-Chinchilla}, T., Vourlidas, A., Raymond, J.~C., {et~al.} 2018, Solar
  Physics, 293

\bibitem[{Parker(1958)}]{parker1958}
Parker, E.~N. 1958, ApJ, 128, 664

\bibitem[{Poedts {et~al.}(2020)Poedts, Andrea, Camilla, Christine, Nicolas,
  Giovanni, Brecht, Dimitrios, Elena, Emmanuel, Tinatin, \&
  Evangelia}]{poedts2020}
Poedts, S., Andrea, L., Camilla, S., {et~al.} 2020, 14

\bibitem[{Regnault {et~al.}(2020)Regnault, Janvier, D{\'e}moulin, Auch{\`e}re,
  Strugarek, Dasso, \& No{\^u}s}]{regnault2020}
Regnault, F., Janvier, M., D{\'e}moulin, P., {et~al.} 2020, J. Geophys. Res.
  Space Physics, 125

\bibitem[{R{\'e}ville \& Brun(2017)}]{reville2017}
R{\'e}ville, V. \& Brun, A.~S. 2017, ApJ, 850, 45

\bibitem[{R{\'e}ville {et~al.}(2015)R{\'e}ville, Brun, Matt, Strugarek, \&
  Pinto}]{reville2015}
R{\'e}ville, V., Brun, A.~S., Matt, S.~P., Strugarek, A., \& Pinto, R.~F. 2015,
  ApJ, 798, 116

\bibitem[{R{\'e}ville {et~al.}(2020)R{\'e}ville, Velli, Panasenco, Tenerani,
  Shi, Badman, Bale, Kasper, Stevens, Korreck, Bonnell, Case, de~Wit, Goetz,
  Harvey, Larson, Livi, Malaspina, MacDowall, Pulupa, \&
  Whittlesey}]{reville2020}
R{\'e}ville, V., Velli, M., Panasenco, O., {et~al.} 2020, ApJS, 246, 24

\bibitem[{Richardson(2004)}]{richardson2004a}
Richardson, I.~G. 2004, Geophysical Research Letters, 31

\bibitem[{Riley \& Crooker(2004)}]{riley2004b}
Riley, P. \& Crooker, N.~U. 2004, ApJ, 600, 1035

\bibitem[{Riley {et~al.}(2004)Riley, Linker, Lionello, Miki{\'c}, Odstrcil,
  Hidalgo, Cid, Hu, Lepping, Lynch, \& Rees}]{riley2004a}
Riley, P., Linker, J., Lionello, R., {et~al.} 2004, Journal of Atmospheric and
  Solar-Terrestrial Physics, 66, 1321

\bibitem[{Ruffenach {et~al.}(2012)Ruffenach, Lavraud, Owens, Sauvaud, Savani,
  Rouillard, D{\'e}moulin, Foullon, Opitz, Fedorov, Jacquey, G{\'e}not, Louarn,
  Luhmann, Russell, Farrugia, \& Galvin}]{ruffenach2012}
Ruffenach, A., Lavraud, B., Owens, M.~J., {et~al.} 2012, Journal of Geophysical
  Research: Space Physics, 117

\bibitem[{Salman {et~al.}(2020)Salman, Winslow, \& Lugaz}]{salman2020a}
Salman, T.~M., Winslow, R.~M., \& Lugaz, N. 2020, Journal of Geophysical
  Research: Space Physics, 125, e2019JA027084

\bibitem[{Schatten {et~al.}(1969)Schatten, Wilcox, \& Ness}]{schatten1969}
Schatten, K.~H., Wilcox, J.~M., \& Ness, N.~F. 1969, Sol Phys, 6, 442

\bibitem[{Schwadron \& McComas(2008)}]{schwadron2008}
Schwadron, N.~A. \& McComas, D.~J. 2008, ApJ, 686, L33

\bibitem[{Scolini {et~al.}(2021)Scolini, Dasso, Rodriguez, Zhukov, \&
  Poedts}]{scolini2021}
Scolini, C., Dasso, S., Rodriguez, L., Zhukov, A.~N., \& Poedts, S. 2021, A\&A,
  649, A69

\bibitem[{Shafranov(1966)}]{shafranov1966}
Shafranov, V.~D. 1966, Reviews of Plasma Physics, 2, 103

\bibitem[{Shen {et~al.}(2014)Shen, Shen, Zhang, Hess, Wang, Feng, Cheng, \&
  Yang}]{shen2014}
Shen, F., Shen, C., Zhang, J., {et~al.} 2014, Journal of Geophysical Research:
  Space Physics, 119, 7128

\bibitem[{Shiota {et~al.}(2010)Shiota, Kusano, Miyoshi, \&
  Shibata}]{shiota2010}
Shiota, D., Kusano, K., Miyoshi, T., \& Shibata, K. 2010, ApJ, 718, 1305

\bibitem[{Siscoe \& Odstrcil(2008)}]{siscoe2008}
Siscoe, G. \& Odstrcil, D. 2008, Journal of Geophysical Research: Space
  Physics, 113

\bibitem[{Sturrock(1966)}]{sturrock1966}
Sturrock, P.~A. 1966, Nature, 211, 695

\bibitem[{Su {et~al.}(2006)Su, Golub, Van~Ballegooijen, \& Gros}]{su2006}
Su, Y.~N., Golub, L., Van~Ballegooijen, A.~A., \& Gros, M. 2006, Sol Phys, 236,
  325

\bibitem[{Titov {et~al.}(2014)Titov, T{\"o}r{\"o}k, Mikic, \&
  Linker}]{titov2014}
Titov, V.~S., T{\"o}r{\"o}k, T., Mikic, Z., \& Linker, J.~A. 2014, The
  Astrophysical Journal, 790, 163

\bibitem[{Toro(2009)}]{toro2009}
Toro, E.~F. 2009, Riemann Solvers and Numerical Methods for Fluid Dynamics: A
  Practical Introduction, 3rd edn. ({Dordrecht ; New York}: {Springer})

\bibitem[{T{\"o}r{\"o}k {et~al.}(2018)T{\"o}r{\"o}k, Downs, Linker, Lionello,
  Titov, Miki{\'c}, Riley, Caplan, \& Wijaya}]{torok2018}
T{\"o}r{\"o}k, T., Downs, C., Linker, J.~A., {et~al.} 2018, The Astrophysical
  Journal, 856, 75

\bibitem[{Verbeke {et~al.}(2019)Verbeke, Pomoell, \& Poedts}]{verbeke2019}
Verbeke, C., Pomoell, J., \& Poedts, S. 2019, Astronomy \& Astrophysics, 627,
  A111

\bibitem[{Wang {et~al.}(2005)Wang, Du, \& Richardson}]{wang2005}
Wang, C., Du, D., \& Richardson, J.~D. 2005, Journal of Geophysical Research:
  Space Physics, 110

\bibitem[{Warren {et~al.}(2011)Warren, O'Brien, \& Sheeley}]{warren2011}
Warren, H.~P., O'Brien, C.~M., \& Sheeley, N.~R. 2011, ApJ, 742, 92

\bibitem[{Winslow {et~al.}(2015)Winslow, Lugaz, Philpott, Schwadron, Farrugia,
  Anderson, \& Smith}]{winslow2015}
Winslow, R.~M., Lugaz, N., Philpott, L.~C., {et~al.} 2015, Journal of
  Geophysical Research: Space Physics, 120, 6101

\bibitem[{Wu \& Lepping(2011)}]{wu2011}
Wu, C.-C. \& Lepping, R.~P. 2011, Solar Physics, 269, 141

\bibitem[{Zhang {et~al.}(2013)Zhang, Hess, \& Poomvises}]{zhang2013}
Zhang, J., Hess, P., \& Poomvises, W. 2013, Solar Physics, 284, 89

\bibitem[{Zwaan(1987)}]{zwaan1987}
Zwaan, C. 1987, Annu. Rev. Astron. Astrophys., 25, 83

\end{thebibliography}

\end{document}